\RequirePackage{rotating}
\documentclass[]{emulateapj}
\pdfoutput=1
\usepackage{times, graphicx, setspace, subfigure, latexsym, amssymb, amsmath, fancyhdr}
\usepackage{natbib}
\usepackage{multirow}
\usepackage{color}

\shorttitle{Monitoring Interstellar Scattering Delays}
\shortauthors{L. Levin et al.}

\begin{document}

\title{The NANOGrav Nine-year Data Set: Monitoring Interstellar Scattering Delays}

\author{
Lina~Levin\altaffilmark{1,2}, Maura~A.~McLaughlin\altaffilmark{1}, Glenn~Jones\altaffilmark{3}, James~M.~Cordes\altaffilmark{4}, Daniel~R.~Stinebring\altaffilmark{5}, Shami~Chatterjee\altaffilmark{4}, Timothy~Dolch\altaffilmark{4,6}, Michael~T.~Lam\altaffilmark{4}, T.~Joseph~W.~Lazio\altaffilmark{7}, Nipuni~Palliyaguru\altaffilmark{1}, Zaven~Arzoumanian\altaffilmark{8}, Kathryn~Crowter\altaffilmark{9}, Paul~B.~Demorest\altaffilmark{10}, Justin~A.~Ellis\altaffilmark{7}, Robert~D.~Ferdman\altaffilmark{11}, Emmanuel~Fonseca\altaffilmark{9}, Marjorie~E.~Gonzalez\altaffilmark{9,12}, Megan~L.~Jones\altaffilmark{1}, David~J.~Nice\altaffilmark{13}, Timothy~T.~Pennucci\altaffilmark{14}, Scott~M.~Ransom\altaffilmark{15}, Ingrid~H.~Stairs\altaffilmark{9,11}, Kevin~Stovall\altaffilmark{16}, Joseph~K.~Swiggum\altaffilmark{1}, Weiwei~Zhu\altaffilmark{9,17}}

\altaffiltext{1}{Department of Physics and Astronomy, West Virginia University, P.O.
Box 6315, Morgantown, WV 26505, USA}
\altaffiltext{2}{Jodrell Bank Centre for Astrophysics, Alan Turing Building, School of Physics and Astronomy, 
The University of Manchester, Oxford Road, Manchester, M13 9PL, UK}
\altaffiltext{3}{Department of Physics, Columbia University, 550 W. 120th St., 
New York, NY 10027, USA}
\altaffiltext{4}{Department of Astronomy, Cornell University, Ithaca, NY
14853, USA}
\altaffiltext{5}{Department of Physics and Astronomy, Oberlin College,
Oberlin, OH 44074, USA}
\altaffiltext{6}{Department of Physics, Hillsdale College, 33 E. College Street, 
Hillsdale, Michigan 49242, USA}
\altaffiltext{7}{Jet Propulsion Laboratory, California Institute of
Technology, 4800 Oak Grove Drive, Pasadena, CA 91106, USA}
\altaffiltext{8}{Center for Research and Exploration in Space Science
and Technology and X-Ray Astrophysics Laboratory, NASA Goddard Space
Flight Center, Code 662, Greenbelt, MD 20771, USA}
\altaffiltext{9}{Department of Physics and Astronomy, University of
British Columbia, 6224 Agricultural Road, Vancouver, BC V6T 1Z1, Canada}
\altaffiltext{10}{National Radio Astronomy Observatory, P. O. Box 0, 
Socorro, NM, 87801, USA}
\altaffiltext{11}{Department of Physics, McGill University, 3600 rue
Universite, Montreal, QC H3A 2T8, Canada}
\altaffiltext{12}{Department of Nuclear Medicine, Vancouver Coastal Health Authority,
 Vancouver, BC V5Z 1M9, Canada}
\altaffiltext{13}{Department of Physics, Lafayette College, Easton, PA
18042, USA}
\altaffiltext{14}{University of Virginia, Department of Astronomy, 
P. O. Box 400325 Charlottesville, VA 22904-4325, USA}
\altaffiltext{15}{National Radio Astronomy Observatory, 520 Edgemont
Road, Charlottesville, VA 22903, USA}
\altaffiltext{16}{Department of Physics and Astronomy, University of New
Mexico, NM 87131, USA}
\altaffiltext{17}{Max-Planck-Institut f{\"u}r Gravitationsphysik, Albert
Einstein Institut, Am M{\"u}lenber 1, 14476 Golm, Germany}



\begin{abstract}
We report on an effort to extract and monitor interstellar scintillation parameters in regular timing observations collected for the NANOGrav pulsar timing array. Scattering delays are measured by creating dynamic spectra for each pulsar and observing epoch of wide-band observations centered near 1500\,MHz and carried out at the Green Bank Telescope and the Arecibo Observatory. The $\sim$800-MHz wide frequency bands imply dramatic changes in scintillation bandwidth across the bandpass, and a stretching routine has been included to account for this scaling. For most of the 10 pulsars for which the scaling has been measured, the bandwidths scale with frequency less steeply than expected for a Kolmogorov medium. We find estimated scattering delay values that vary with time by up to an order of magnitude. 
The mean measured scattering delays are similar to previously published values and slightly higher than predicted by interstellar medium models.
We investigate the possibility of increasing the timing precision by mitigating timing errors introduced by the scattering delays. For most of the pulsars, the uncertainty in the time of arrival of a single timing point is much larger than the maximum variation of the scattering delay, suggesting that diffractive scintillation remains only a negligible part of their noise budget. 
\end{abstract}


\section{Introduction}
The interstellar medium (ISM) consists partly of ionized plasma, which interacts with pulsar radio emission. This has several effects on pulsar signals, and will impact the times of arrival (TOAs) of the pulses at Earth. 
One such effect is dispersion, which occurs when the radio wave propagates through a column of free electrons in the ISM and is characterized by a frequency-dependent time delay. The delays are proportional to DM $\times$ $\nu^{-2}$, where the dispersion measure, DM, is the integrated column density of free electrons and $\nu$ is the observing frequency. 
Since the pulsar and the ISM have different relative velocities, the DM of a pulsar is not constant in time. By observing pulsars at two or more separate frequencies, the DM variations can be tracked and corrected for in the data \citep[e.g.,][]{dem13, kei13, lee14}. 

In contrast to dispersion, which would be present in a completely homogeneous medium, scattering arises when radio waves travel through an inhomogeneous medium. Multi-path scattering manifests itself in several ways, including diffractive intensity scintillations and pulse broadening. Diffractive scintillation effects were first observed in pulsars by \cite{lyn68}, and are the effects in focus in this paper. 
The basic model usually used to describe diffractive scintillation assumes that the ISM is a thin screen of plasma, located between the pulsar and the observer \citep{sch68}. As the signal propagates through the screen, inhomogeneities in the plasma introduce phase perturbations that are correlated over a scintillation bandwidth, which is inversely proportional to the scattering timescale. These perturbations are also known as scattering delays. 
The scintillation pattern, and hence the scattering timescale, of a pulsar can change drastically over time \citep{hem08}. Similar to the case of DM variations, the relative velocities of the pulsar and the ISM give rise to the time variable scattering delays. 
The scaling of the scintillation parameters with frequency is often described as that of a Kolmogorov medium, with the scintillation bandwidth, $\Delta \nu_{\rm d} \propto \nu^{4.4}$, and the scintillation timescale, $\Delta t_{\rm d} \propto \nu^{1.2}$ \citep{cor85}, although it has been shown that some sources have scaling indices that deviate from these \citep[e.g.,][]{loh04, bha04}. These observed deviations may not necessarily be indicative of a non-Kolmogorov spectrum, but may be due to the size of the dominant scattering region transverse to the line of sight \citep{cor01}. 

In a Pulsar Timing Array (PTA), millisecond pulsars (MSPs) are observed in an effort to detect nanohertz gravitational waves. The North American Nanohertz Observatory for Gravitational Waves (NANOGrav) uses the 100-m Green Bank Telescope (GBT) and the 300-m Arecibo Observatory (AO) to observe $\sim$40 MSPs every 7$-$28 days. To succeed in detecting gravitational waves, it is necessary to obtain as high a timing precision as possible, over a long time span. 
The achievable timing precision of MSPs is continually increasing with longer data sets and improved instrumentation. This makes it ever more important to understand all non-gravitational wave effects, to improve the sensitivity of PTAs to gravitational waves. 

As described above, the perturbations caused by interstellar scintillation, as well as the perturbations from DM variations, limit the timing precision for all pulsars. Some pulsars will be more affected than others, however, depending on the properties of the ISM along their line of sight. 
Most MSPs in PTAs have been chosen partly due to having a low DM, and hence it is expected that scattering will only contribute to a small part of the total timing error for these pulsars. 
However, accurately correcting for ISM perturbations needs to be done carefully, so as not to introduce additional errors. To correct for DM variations, nearly simultaneous observations with at least two separate frequency bands at widely spaced center frequencies are used. Because of the different frequency scaling of dispersion and interstellar scattering, by only correcting for DM variations systematic errors are introduced into the timing procedure (see Appendix \ref{app:contamination}).
One important caveat to this correction procedure is that measurements at different observing frequencies sample different parts of the interstellar medium \citep{cor15}. This significantly affects DM measurements and could also affect interstellar scintillation measurements for some pulsars.

A number of authors have addressed the issues of measuring scintillation parameters and mitigating ISM effects previously.  
A few examples include \cite{col10} and \cite{kei13}, who analyzed DM variations and scintillation parameters for pulsars in the Parkes Pulsar Timing Array (PPTA). Their observations were carried out at three different frequencies: a 64-MHz band centered at 685 MHz, a 256-MHz band centered at 1369 MHz, and a 1024-MHz band centered at 3100 MHz \citep{man13}. Some of the pulsars in the PPTA sample are also observed by NANOGrav and a comparison of the results is included in this work. 
In a different paper, \cite{gup94} studied the scintillation properties of 8 pulsars over a 16 month period with the Lovell telescope at Jodrell Bank, using a 5-MHz band centered at 408\,MHz. They found that the observed fluctuations in the spectra could be explained as refractive modulation of the diffractive scintillation parameters.
\cite{bha98} analyzed scintillation parameters for 20 slow pulsars with low DM ($<$\,35\,pc\,cm$^{-3}$), using observations at a 9-MHz frequency band centered at 327\,MHz at 10$-$90 epochs spanning $\sim$100\,days. They report large fluctuations in both diffractive scintillation timescale and bandwidth, with variations of a factor of 3$-$5 for most pulsars in their sample. 

In this paper, we analyze the magnitude and variation of interstellar scattering delays in regular NANOGrav timing observations to investigate the effect of interstellar scintillation and its contribution to the total noise budget for this important set of pulsars.

\section{Data}
This paper makes use of data from regular NANOGrav timing observations. The data are all included in the latest NANOGrav data release \citep[9-year dataset;][]{arz15}, and here we use a sub-set spanning $\sim$3.7\,years for data from GBT and $\sim$1.7\,years for data from AO. The exact MJD range of the data used for each pulsar is shown in Table\,\ref{tab:allscatt}, together with observational properties of the pulsar as well as of the ISM along the line-of-sight to the pulsar.
We have focused on observations carried out at a center frequency of $\sim$1500\,MHz for 20 pulsars at GBT and 19 pulsars at AO. Two of the pulsars (J1713+0747 and B1937+21) are observed with both telescopes. 

At GBT, the data are collected with the FPGA-based spectrometer GUPPI (Green Bank Ultimate Pulsar Processing Instrument) using coherent dedispersion techniques. The observations are carried out over a frequency band of 800\,MHz centered at 1500\,MHz, divided into 1.5625-MHz wide frequency channels. 
All observations are $\sim$30 minutes in length and are folded in real time with 15-s subintegrations.

At AO, we are using data collected and coherently dedispersed with the PUPPI (Puerto Rico Ultimate Pulsar Processing Instrument) backend. Here the observations are conducted over a 700-MHz bandwidth centered near 1500\,MHz, divided in 1.5625-MHz wide channels. 
The data are recorded in 1-s subintegrations (or 10-s subintegrations for observations before MJD$\sim$56540) for $\sim$30 minutes per pulsar and epoch. 

At the start of each observation, a polarization calibration scan is performed by injecting a 25-Hz noise diode for both polarizations. Once during each epoch and for each observing frequency, a flux calibrator (B1442+101) is observed. For the analysis in this paper, total intensity profiles have been used, by summing the polarizations of the calibrated data \citep{arz15}.  

The GUPPI and PUPPI backends provide substantially larger observation bandwidths compared to previously used backends ASP (Astronomical Signal Processor) and GASP (Green Bank Astronomical Signal Processor), which both had 64-MHz bandwidth capacity \citep[e.g.,][]{dem13}. The wider bandwidths not only result in higher timing precision due to a higher signal-to-noise value for the pulsar signal overall, but they also provide a larger number of scintillation maxima and minima over the observed band. The wide-band observations carried out with GUPPI and PUPPI prompted a need to investigate the effect of interstellar scattering delays, and are essential for the analysis in this paper.

\section{Analysis}
\label{sec:analysis}
\begin{figure}
	\begin{center}
		\includegraphics[width=0.45\textwidth]{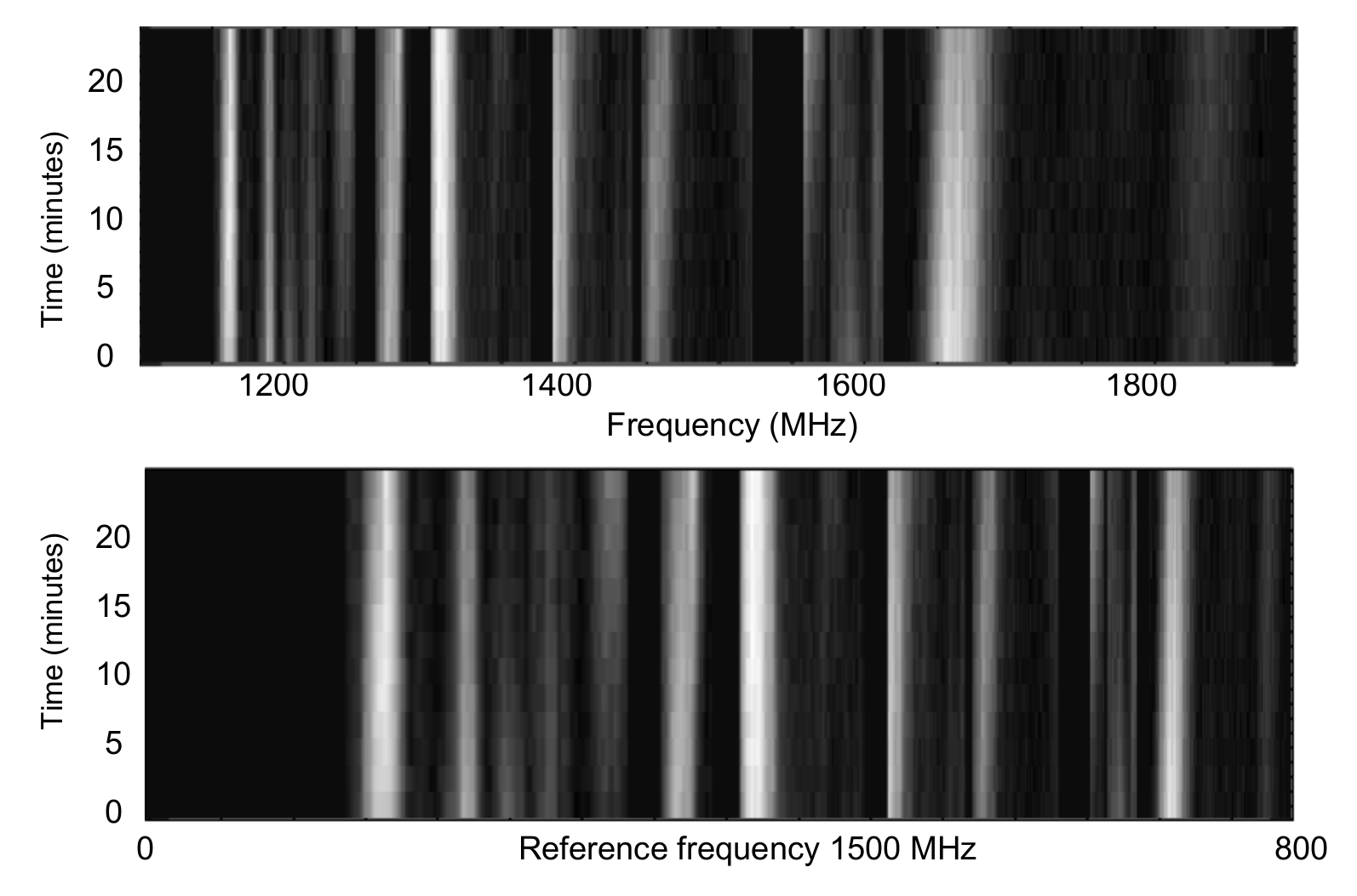}
	\caption{The top panel shows an original dynamic spectrum for PSR\,J1918--0642 at MJD 56066. The bottom panel shows the same spectrum after it has been stretched to a reference frequency of 1500\,MHz, using a scaling index of $\zeta_{\rm stretch} = 4.4$.}
	\label{fig:dynspec}
	\end{center}
\end{figure}

We have created and analyzed 2-dimensional dynamic spectra of each 1500-MHz observation for each of the NANOGrav pulsars, following a procedure similar to that described in \cite{cor86b} unless otherwise stated (e.g. for part of the delay uncertainties as described in eq \ref{eq:finitescintle} below). A dynamic spectrum displays how the intensity of the pulsar signal varies with time, $t$, and observing frequency, $\nu$. Each point in the spectrum is calculated by 
\begin{equation}
	S(\nu, t) = \frac{P_{\rm on}(\nu, t) - P_{\rm off}(\nu, t)}{P_{\rm bandpass}(\nu)}
\end{equation}
where $P_{\rm bandpass}$ is the total power of the observation as a function of observing frequency, and $P_{\rm on}$ and $P_{\rm off}$ are the power in the on- and off-pulse part of the pulse profile respectively. 
The on-pulse part is here defined as all bins in the summed pulse profile that have an intensity $> 5\%$ of the maximum intensity, after smoothing the original 2048 pulse profile bins down to 64 bins. This is done for each observation of each pulsar. 
An example of a dynamic spectrum can be found in the top panel of Fig.\,\ref{fig:dynspec}, where lighter pixels indicate greater interference between radio waves and each local maximum is known as a scintle.
To calculate the sizes of the scintles and hence analyze the interference pattern, we compute a 2D autocorrelation function (ACF). The ACF is then summed over time and frequency separately and a Gaussian function, centered at zero lag, is fitted to the two 1D ACFs to determine the scintillation parameters. The scintillation timescale, $\Delta t_{\rm d}$, is defined as the half-width at $e^{-1}$ of the summed frequency lag and the scintillation bandwidth, $\Delta \nu_{\rm d}$, is the half-width at half-maximum of the summed time lag. Most of the observations have total integration time $T < \Delta t_{\rm d}$ at our observing frequencies, 
and hence we cannot calculate values for $\Delta t_{\rm d}$. 

The scattering delay, $\tau_{\rm d}$, that arises as a result of scintillation can be calculated from
\begin{equation}
	2\pi \Delta \nu_{\rm d} \tau_{\rm d} = C_{\rm 1} 
	\label{eq:delay}
\end{equation}
where $C_{\rm 1}$ is a constant with value ranging 0.6$-$1.5 \citep{lam99} depending on the geometry and spectrum of the electron density fluctuations. Here we assume $C_{\rm 1} = 1$.

The uncertainties of the scattering delay measurements consist partly of a ``finite scintle error" and partly of the uncertainty of the least-square fit of a Gaussian to the ACF, which are added in quadrature to get the total uncertainty. The finite scintle error is calculated as
\begin{eqnarray}
	\epsilon & \approx & \tau_{\rm d} N_{\rm scint}^{-1/2} \nonumber \\ 
	              & \approx & \tau_{\rm d} \left[ (1 + \eta_{\rm t}T/\Delta t_{\rm d}) (1 + \eta_{\nu}B/\Delta \nu_{\rm d}) \right]^{-1/2}
	\label{eq:finitescintle}
\end{eqnarray}
where $N_{\rm scint}$ is the number of scintles, $T$ and $B$ are total integration time and total bandwidth, respectively, and $\eta_{\rm t}$ and $\eta_{\nu}$ are filling factors in the range $0.1-0.3$, here set to 0.2 \citep{cor10}. Since commonly for our observations $T \ll \Delta t_{\rm d}$, the first term in Equation \ref{eq:finitescintle} is approximately equal to 1 and hence $\epsilon$ depends only on values related to the bandwidth.
\footnote{If $T \approx \Delta t_{\rm d}$, implying that the first term would be equal to 1.2, we would be slightly underestimating $N_{\rm scint}$ and hence slightly overestimating the scattering delay uncertainties.}

A positive consequence of the wide bandwidth observations is the larger numbers of scintles observed compared to observations with narrower bandwidths. However, it also implies large differences in the scintillation bandwidths measured at the lowest part of the band compared with the higher part of the band, which in turn causes problems when creating ACFs. 
This is only problematic if the collected data span a large range in frequency. Current methods to calculate scintillation parameters were developed for narrower bands, and do not provide a direct solution to the scaling issue.
To resolve this, we have developed a method to ``stretch" the spectra to a reference frequency based on how the scintillation bandwidth scales with observing frequency. 

\begin{figure}
	\begin{center}
		\includegraphics[width=0.45\textwidth]{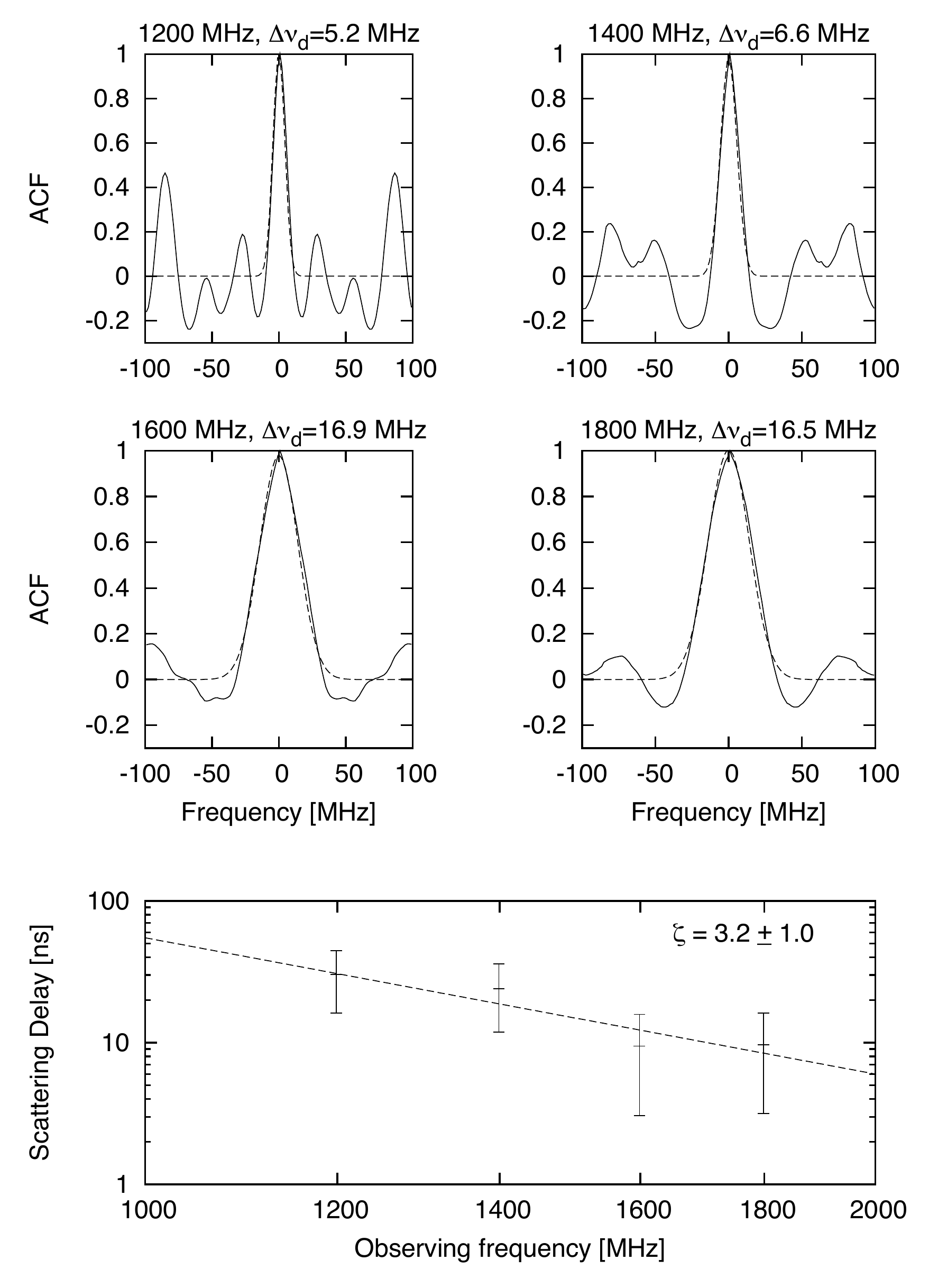}
	\caption{PSR\,J1918--0642 scattering delay scalings over the frequency band for an observation from MJD 56066. The four top panels show the autocorrelation function of each 200-MHz subband as a solid line and the Gaussian fit as a dashed line. The best-fit value for each scintillation bandwidth, $\Delta \nu_{\rm d}$, is converted into a scattering delay and plotted in the bottom panel, together with the best fit of the scaling index, $\zeta$, as a dashed line.}
	\label{fig:1918scattscale}
	\end{center}
\end{figure}

To investigate the scaling, we divided the wide bandwidth observations into four equally sized sub-bands. The scattering delays inferred from the measured scintillation bandwidths of the four separate bands were then plotted against the center frequency of the bands, and a function of the form $\tau_{\rm d} = \nu^{-\zeta}$ was fitted to the data to calculate the scaling index, $\zeta$. An example of a scaling index measurement is given in Fig.\,\ref{fig:1918scattscale}.
Noting that the number of scintles in each sub-band drastically changes with observing frequency, which could potentially affect this analysis, we also tried dividing the band up into four parts with the same number of scintles per band instead of a set bandwidth. The result of this exercise agreed well with the set bandwidth method, and hence the values reported here are all calculated from 200-MHz sub-bands. 

This method was used to measure scaling indices for 10 of the NANOGrav pulsars. The remaining pulsars all have either too wide scintillation bandwidths to get reliable measurements in a 200-MHz band, or too narrow scintles to get resolved scintillation bandwidths with the 1.5625-MHz frequency channels. 
However, in an effort to keep the number of variables to a minimum, we choose to use a scaling index as predicted in a Kolmogorov medium ($\zeta_{\rm stretch}$ = 4.4) in the stretching even for the sources with a measured $\zeta$-value. 
This choice is also based on the observation of possible variable scaling indices between epochs (see Sec \ref{sec:scalingindex}), and hence using a measured scaling index from one epoch in the stretching of another epoch may bias the dynamic spectrum analysis. 
In addition, comparisons show that the consequences of using a fixed $\zeta_{\rm stretch}$ are smaller than the other statistical uncertainties on the scattering delay measurements. 

Hence, each dynamic spectrum was stretched using a scaling index $\zeta_{\rm stretch}$ = 4.4, by rescaling the frequency axis and setting the reference frequency, $\nu_{\rm ref}$, to the center of the band. The result is an 800-MHz wide dynamic spectrum with evenly-sized scintles that all refer to the same observing frequency. 
We calculated ACFs of these stretched spectra and derived from them the scintillation bandwidths that are used in the remainder of this paper. 
See the bottom panel of Fig.\,\ref{fig:dynspec} for an example of a stretched spectrum.

\section{Results}
\begin{sidewaystable*}
  \begin{center}
\vspace{7cm}
  \caption{Scattering properties of NANOGrav MSPs.}
  \label{tab:allscatt}
  \begin{tabular}{l l l l l l l l l l l l l l l}
	\hline 
	\hline
	Pulsar & period    & DM          & S$_{\rm 1400}$        & T$^{\rm NE2001}_{\rm ref}$ & $\tau^{\rm NE2001}_{\rm d}$ &  $\bar{\tau}_{\rm d}$  & $\Delta \bar{\nu_{\rm d}}$ & SM & N$_{\rm scint}^{\rm med}$ &  N$_{\rm m}$ & N$_{\rm obs}$ & $\Delta \tau_{\rm d}$ & ${\sigma}_{\rm \small TOA}^{\rm med}$ & MJD range\\
		   & {\small(ms)}       & {\small(pc\,cm$^{-3}$)} & {\small(mJy)} &    {\small(days)}		           &  {\small(ns)}         & {\small(ns)}    &  {\small(MHz)}    & {\small(kpc m$^{-20/3}$)} & & & & {\small(ns) }                          & {\small(ns)}                           &  {\small(days)} \\
	\hline	
J0023+0923   &    3.05   &   14.3   &  0.48            	&   0.28     & 2.5         &   6.2 $\pm$ 2.1  & 20.5 $\pm$ 6.6 & 2.9 $\times 10^{-4}$ & 7 & 5 & 24 & 6.3              & 72  & 55989--56599 \\
J0030+0451   &    4.87   &    4.3   & 1.1    			&   0.029   & 0.055     &   --                     & --                       & -- 				& -- & --  & 27 & --                 & 104  & 55989--56599 \\
J0340+4130   &    3.29   &   49.6   & 0.31           	&    2.0     & 49      &   14.7 $\pm$ 5.5    & 9.1 $\pm$ 3.3  & 2.6 $\times 10^{-4}$ & 16 & 15  &  28 & 23.6       & 638 & 55972--56586 \\
J0613--0200   &    3.06   &   38.8   & 1.7       		&    1.1     & 16      &   11.7 $\pm$ 3.9    & 11.1 $\pm$ 4.3  & 2.2 $\times 10^{-4}$ & 12 & 25   & 49 & 20.7     & 180 & 55275--56586 \\
J0645+5158   &    8.85   &   18.2   &  0.29                	&    0.45   & 6.2     &   --                        &  --                        &  -- 			      & --  &    --   &  33   & --          & 394  & 55700--56586 \\
 & & & & & & & & & & & & &\\
J0931--1902   &    4.64   &   41.5   &  0.42                 	&  1.5       & 24     &    3.2                     & 50.3                     & 6.4 $\times 10^{-3}$  &  4 & 1    &  11  &    --     & 622 & 56351--56586 \\
J1012+5307   &    5.26   &    9.0   & 3.2        		&    0.15   & 1.2     &   2.5 $\pm$ 0.1     & 66.4 $\pm$ 5.6   &  1.8 $\times 10^{-4}$ & 3 & 4   &  51 & 0.5        & 207 & 55275--56586 \\
J1024--0719   &    5.16   &    6.5   &  1.5             	&    0.055  & 0.17  &   2.8 $\pm$ 1.3     & 46.8 $\pm$ 17.8  &  2.4 $\times 10^{-4}$   & 7 & 5   &  52 & 3.8     & 338 & 55275--56586 \\
J1455--3330   &    7.99   &   13.6   & 0.65       		&    0.15   & 0.94   &   4.0 $\pm$ 1.1     & 69.6 $\pm$ 17.5   &  1.6 $\times 10^{-4}$  & 3 & 4   &  51  & 3.1    & 706 & 55275--56584 \\
J1600--3053   &    3.60   &   52.3   & 2.2       		&    2.7     & 93      &   --                        &  --                         &   --   			& -- &   --    &  50 & --          &  81  & 55275--56584 \\
 & & & & & & & & & & & & &\\
J1614--2230   &    3.15   &   34.5   & 0.70                 	&    1.3     & 29      &   16.2 $\pm$ 4.3   &  9.0 $\pm$ 2.6    & 3.5 $\times 10^{-4}$  & 18 &  42   &  64   & 18.1   & 170 & 55265--56584 \\
J1640+2224   &    3.16   &   18.4  & 0.70    		&    0.56    & 5.8    &   2.6 $\pm$ 1.1     &  56.3 $\pm$ 14.6 & 8.8 $\times 10^{-3}$  & 4   &  9  &  32 &  2.8        & 33 & 56023--56598 \\
J1643--1224   &    4.62   &   62.4   & 3.8      		&    3.2     & 90      &   --                        &  --                         &   --   		             & --   &  --      &  51 & --       & 192 & 55275--56584 \\
J1713+0747   &    4.57   &   16.0   & 8.5      		&    0.42   & 4.1     &   7.1 $\pm$ 2.4    & 21.1 $\pm$ 8.6     &  2.3 $\times 10^{-4}$ & 8   &  62   &  110  & 11.9  & 20 & 55275--56598 \\
J1738+0333   &    5.85   &   33.8  & 0.67               	&    1.2     & 21      &    9.4 $\pm$ 2.3   &  16.8 $\pm$ 7.7  & 1.9 $\times 10^{-4}$  & 11  &  12   &  20 & 11.1     & 126 & 56018--56591 \\
 & & & & & & & & & & & & &\\
J1741+1351   &    3.75   &   24.0   & 0.37            	&    0.19   & 0.86      &   5.0 $\pm$ 4.4   &  17.1 $\pm$ 17.3  &  2.2 $\times 10^{-4}$ & 4 & 10   &  24 & 20.7     & 93  & 55996--56593 \\
J1744--1134   &    4.08   &    3.1   &  2.8      		&    0.020  & 0.020   &   3.8 $\pm$ 1.3   &  42.1 $\pm$ 9.1 & 2.7 $\times 10^{-4}$  & 5 &  12    &  47  & 4.1       & 81  & 55275--56584 \\
J1747--4036   &    1.65   &   152.9  & 1.0                 	&   19       & 2\,300   &   --                      &  --                         &  --  				& -- &   --     &  25  & --    & 445 & 55977--56584\\
J1832--0836    &     2.72   &   28.2   & 0.79           	&  0.97     & 18         &    --                     &  --                         &   --   				& -- &   --     &  10   &  --  & 169  & 56367--56584 \\
J1853+1303   &    4.09   &   30.6   & 0.45   		&    0.72    & 5.3       &  11.7 $\pm$ 4.5  &  12.7 $\pm$ 5.1   & 1.7 $\times 10^{-4}$ & 11 &   4   &  24  & 12.0    & 148  & 55989--56598 \\
 & & & & & & & & & & & & &\\ 
B1855+09     &    5.36   &   13.3   & 3.8     			&    0.47   & 4.0          &  21.3 $\pm$ 9.9   & 5.2 $\pm$ 2.4     & 5.4 $\times 10^{-4}$ & 26 &  26   &  33   & 45.3    & 59 & 55989--56598 \\
J1903+0327   &    2.15   &   297.5 &  0.61             	&    260   & 230\,000  &   --                         &    --                      &  --   				& -- &   --   &  21 & --           & 178 & 55997--56591\\
J1909--3744   &    2.95   &   10.4   &  1.6    		&    0.18   & 1.5          &   4.9 $\pm$ 1.8     & 39.0 $\pm$ 14.7  &  2.6 $\times 10^{-4}$  &  5 &  17   &  59  & 7.0    & 41  & 55275--56598 \\
J1910+1256   &    4.98   &   38.1   & 0.50    		&    0.96    & 8.4         &  57.8 $\pm$ 16.6  & 2.3 $\pm$ 0.9      & 6.3 $\times 10^{-4}$ & 63 & 24   &  25 & 83.0     & 96 & 55989--56598 \\
J1918--0642   &    7.65   &   26.6   & 1.4     		&    0.78    & 10          &  9.7 $\pm$ 3.0    & 14.9 $\pm$ 5.1    & 2.3 $\times 10^{-4}$ & 11 &  32    &  49  & 11.8    & 219 & 55275--56584 \\
 & & & & & & & & & & & & &\\
J1923+2515   &    3.88   &   18.9   & 0.22               	&    0.36    & 1.7          &   6.1 $\pm$ 0.8      &  21.6 $\pm$ 9.5  &  1.1 $\times 10^{-4}$  & 6   & 4   &  23 & 3.6          & 214 & 55996--56593 \\
B1937+21     &    1.56   &   71.0   &  13.3               	&    4.7      & 130         &  44.3 $\pm$ 21.4   &  2.8 $\pm$ 1.3    &  3.6 $\times 10^{-4}$  & 43 &  11   &  67  & 88.5    & 5  & 55275--56593 \\
J1944+0907   &    5.19   &   24.3   & 2.6                	&     0.50   & 2.9          &   10.1 $\pm$ 5.6    &  10.6 $\pm$ 5.1  & 2.1 $\times 10^{-4}$   & 12 &  16   &  21 & 23.2    & 161 & 55997--56591 \\
J1949+3106   &   13.14   &   164.1 &  0.12          	&   14       & 610          &   --                         &  --                        &   --   		            & --  & --   &  17 & --           & 566  & 56139--56593 \\
B1953+29     &    6.13   &   104.5  &  0.82   		&     7.3    & 240          &   55.3                    & 2.9                       & 3.2 $\times 10^{-4}$ & 63   & 3   &  21 & --           & 223 & 56018--56591 \\
  \end{tabular}
  \end{center}
\end{sidewaystable*}

\begin{sidewaystable*}
  \begin{center}
  {\small{\sc Scattering properties of NANOGrav MSPs {\it cont.}}}
  \label{tab:allscatt2}
  \begin{tabular}{l l l l l l l l l l l l l l l l}
	\hline 
	\hline
	Pulsar & period    & DM          & S$_{\rm 1400}$        & T$^{\rm NE2001}_{\rm ref}$ & $\tau^{\rm NE2001}_{\rm d}$ &  $\bar{\tau}_{\rm d}$ & $\Delta \bar{\nu_{\rm d}}$  & SM & N$_{\rm scint}^{\rm med}$ & N$_{\rm m}$ & N$_{\rm obs}$ & $\Delta \tau_{\rm d}$ & ${\sigma}_{\rm \small TOA}^{\rm med}$ & MJD range\\
		   & {\small(ms)}       & {\small(pc\,cm$^{-3}$)} & {\small(mJy)} &    {\small(days)}		           &  {\small(ns)}         & {\small(ns)}   & {\small(MHz)}     & {\small(kpc m$^{-20/3}$)} & & & & {\small(ns)}                           & {\small(ns)}                            & {\small(days)} \\
	\hline	 
J2010--1323   &    5.22   &   22.2   & 0.60               	&    0.57    & 6.6      &   19.3 $\pm$ 6.4  & 6.9 $\pm$ 2.3      &  5.1 $\times 10^{-4}$  & 21 & 30   &  48  & 26.5      & 293 & 55275--56584\\
J2017+0603   &    2.90   &   23.9   &  0.28             	&    0.53   & 3.8       &   8.4 $\pm$ 3.4    & 21.6 $\pm$ 9.5  & 1.4 $\times 10^{-4}$  &  8 &  5  &  29 & 8.7               & 95 & 55989--56598 \\
J2043+1711   &    2.38   &   20.7   & 0.094                	&    0.41   & 2.0       &   1.8                     & 86.3                     &  4.2  $\times 10^{-3}$  &  3 &  1   &  33 & --            & 68 & 55997--56591 \\
J2145--0750   &   16.05   &    9.0   & 5.5      		&    0.12    & 0.53    &   2.8 $\pm$ 0.7    & 47.8 $\pm$ 13.3  &  1.7  $\times 10^{-4}$ & 4 & 9    &  48   & 3.2         & 214 & 55275--56584 \\
J2214+3000   &    3.12   &   22.6   &  0.53              	&    0.48   & 3.2       &    3.1 $\pm$ 3.4   & 23.0 $\pm$ 17.9  &  1.2  $\times 10^{-4}$& 6 & 7   &  25 & 10.1           & 160 & 55989--56598 \\
 & & & & & & & & & & & & & &\\
J2302+4442   &    5.19   &   13.8   &  0.90               	&    0.22   & 0.84    &   14.1 $\pm$ 2.7   & 9.9 $\pm$ 2.4   &  3.5 $\times 10^{-4}$ & 15 & 10   &  26   & 12.2       & 637 & 55972--56586 \\
J2317+1439   &    3.45   &   21.9   &  0.80              	&    0.27   & 1.9      &    3.0 $\pm$ 1.0    &  41.8 $\pm$ 11.8 &  1.4 $\times 10^{-4}$ & 6 & 3   &  19 & 2.1            & 81 & 56100--56599 \\
	\hline
	\hline
  \end{tabular}
  \end{center}
	{\small {\bf Notes.} Flux density values at 1400\,MHz (S$_{\rm 1400}$) are average values from the calibrated timing observations. Scattering delays from
	the NE2001 model ($\tau^{\rm NE2001}_{\rm d}$) have been scaled to an observing frequency of 1500\,MHz, and the refractive time scales (T$^{\rm NE2001}_{\rm ref}$)
	are calculated from the scaled values of the scintillation bandwidth and timescale estimated within the model. The weighted average of the scattering delay is given as $\bar{\tau}_{\rm d}$ 
	and the weighted average of the measured scintillation bandwidth is given as $\Delta \bar{\nu_{\rm d}}$, where the error is represented by the standard deviation of the weighted mean in both cases.
	The maximum variation of the scattering delay is calculated as the maximum measured scattering delay minus the minimum measured scattering delay and is here given as $\Delta \tau_{\rm d}$. 
	The scattering measure (SM) is calculated through eq \ref{eq:SM}, using distances estimated from the NE2001 model. 
	The number of epochs included in $\bar{\tau}_{\rm d}$ is given as N$_{\rm m}$, while the total number of epochs analysed is given as N$_{\rm obs}$. 
	The number of scintles in each of the N$_{\rm m}$ epochs are calculated through the expression for N$_{\rm scint}$ given in Equation \ref{eq:finitescintle} and the median number of scintles is given as N$_{\rm scint}^{\rm med}$. 
	The median TOA uncertainty (${\sigma}_{\rm TOA}^{\rm med}$) is given as the averaged residual error for the combined band, with values from \cite{arz15}.}
\end{sidewaystable*}

\begin{table}
  \begin{center}
  \caption{Measured scattering delay scaling indices.}
  \label{tab:scalingindex}
  \begin{tabular}{l l l l}
	\hline 
	\hline
	Pulsar & $\zeta$ & MJD & N$_{\rm scint}^{\rm subband}$ \\
	\hline
	J0613--0200 & 2.8(2) & 55275 & 11 \\
			    & 1.2(8)  & 56130 & 8 \\
	J1614--2230 & 2.6(9)  & 55269 & 6 \\
			    & 6.3(4)  & 55304 & 9 \\
			    & 2.5(4)  & 55892 & 13 \\
			    & 2(2)  & 56288 & 13 \\
	J1713+0747 & 1.1(5) & 55949 & 5 \\ 
			    &  4.1(2) & 56299 & 4 \\
			    & 3.7(5)  & 56391 & 3 \\
	B1855+09    & 3.8(3)  & 56294 & 10 \\
	J1910+1256 & 2.3(6)  & 56319 & 29 \\
	J1918--0642 & 2.4(6)  & 55463 & 5 \\
			     & 3(1)  & 56066 & 5 \\
	B1937+21    & 1.3(3) & 56131 & 11 \\
			    & 2.7(5)  &  56403 & 13 \\
			    &  3(1) & 56451 & 10 \\
			    & 3(1)   & 56492 & 24 \\
			    & 1.8(1) & 56548 & 8 \\
			    & 4.9(8)  &  56593 & 26 \\
	J1944+0907 & 3(2)     & 56018 & 7 \\
	J2010--1323 & 3.9(4) & 56095 & 10 \\
			    &  4.4(1) & 56250 & 10 \\
			    & 1.9(4) & 56352 & 9 \\
			    & 1(1)  & 56472 & 9  \\
			    & 3.1(5)  & 56523 & 23 \\
	J2145--0750 &  3.1(2)  & 56195 & 6 \\
	\hline
	\hline 
  \end{tabular}
  \end{center}
  {\small {\bf Notes.} Scattering delay scaling over frequency band, where $\tau_d \propto \nu^{-\zeta}$. The indices are measured at the given observing epoch, and the numbers in parenthesis are the errors on the last given digit. N$_{\rm scint}^{\rm subband}$ represents the average number of scintles in each subband, calculated through the expression given in equation \ref{eq:finitescintle}.}
\end{table}

\subsection{Scaling over the observed frequency band}
Using the sub-banding method described in Sec.\,\ref{sec:analysis}, we have successfully measured scaling indices for 10 of the 37 NANOGrav MSPs. These values are given in Table \ref{tab:scalingindex}.  
For most of the pulsars, the scaling index is lower than that for a Kolmogorov medium ($\zeta = 4.4$). 
In addition to the variation between different pulsars, the scaling index for a particular pulsar may not necessarily be constant in time 
and ideally also this variation should be monitored. However, it has proven difficult to measure scaling indices at most epochs because scintles are not visible in all of the sub-bands, or because radio frequency interference is severely affecting parts of the band.  
Only a few of the pulsars have multiple scaling index measurements (for further discussion, see Section \ref{sec:scalingindex}).

\subsection{Scattering delay variation}
Interstellar scattering delays ($\bar{\tau}_{\rm d}$) for all NANOGrav pulsars are given as the weighted average over all observing epochs in Table \ref{tab:allscatt}. For comparison, values for the predicted scattering delay and refractive timescales given by the NE2001 model for free electrons in the ISM \citep{cor02} are listed for all the pulsars. 
In addition, the table includes the maximum variation measured ($\Delta \tau_{\rm d} = \tau_{\rm d;max} - \tau_{\rm d;min}$), together with the median TOA uncertainty resulting from timing. Comparing these values can give insight into the feasibility of correcting the timing residuals for scattering delays for a particular pulsar (see Sec. \ref{sec:timing} for more details). 

\begin{figure}
	\begin{center}
		\includegraphics[width=0.45\textwidth]{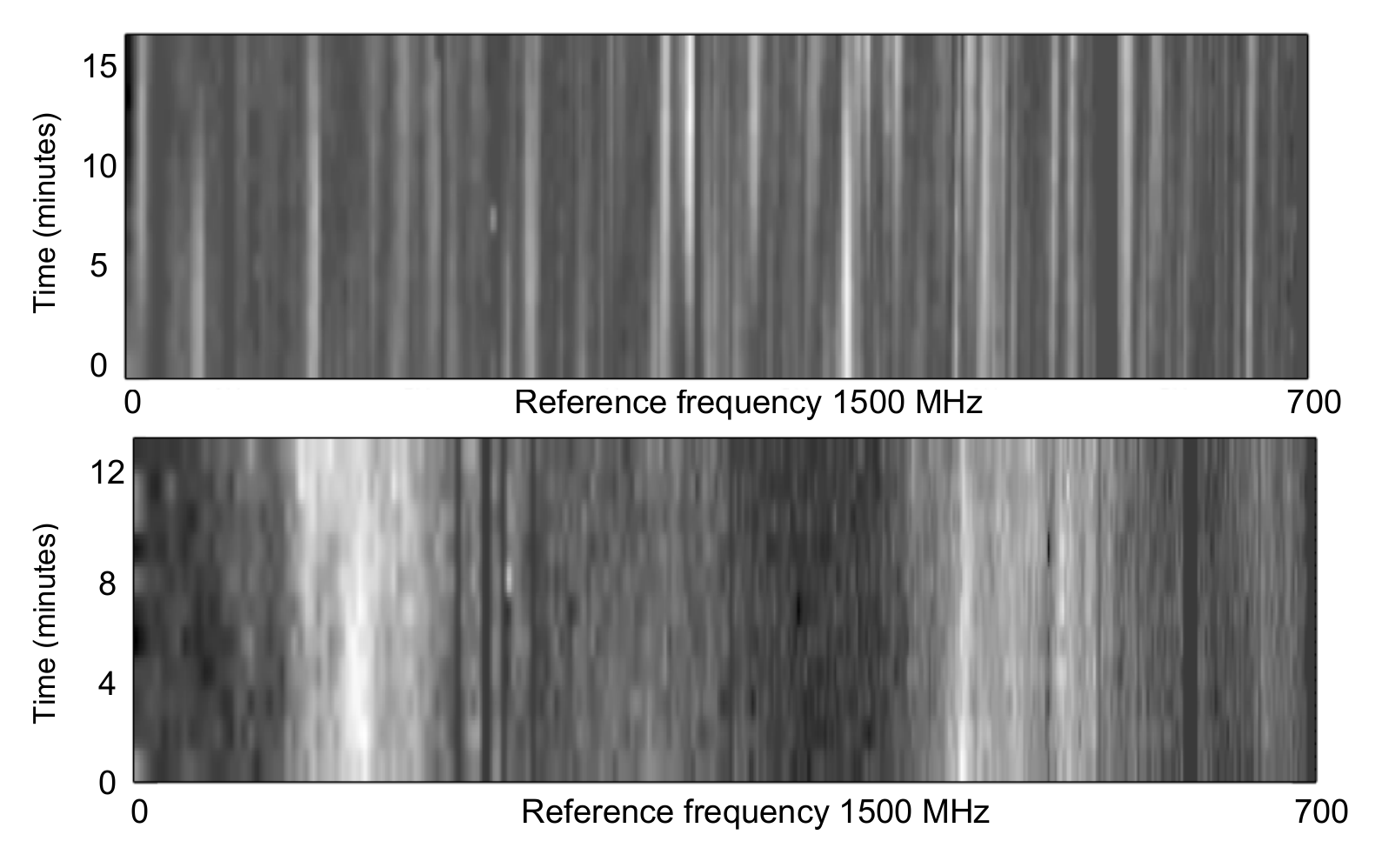}
	\caption{Examples of dynamic spectra with narrow and wide scintillation bandwidths. Both observations are collected with the PUPPI backend at Arecibo, and have been stretched to a reference frequency of 1500\,MHz. The top panel shows an observation of PSR\,J1910+1256 at MJD 56519, and the bottom panel is an observation of J2317+1439 at MJD 56100.}
	\label{fig:dynspecex}
	\end{center}
\end{figure}

\begin{figure*}
	\begin{center}
		\includegraphics[width=0.32\textwidth]{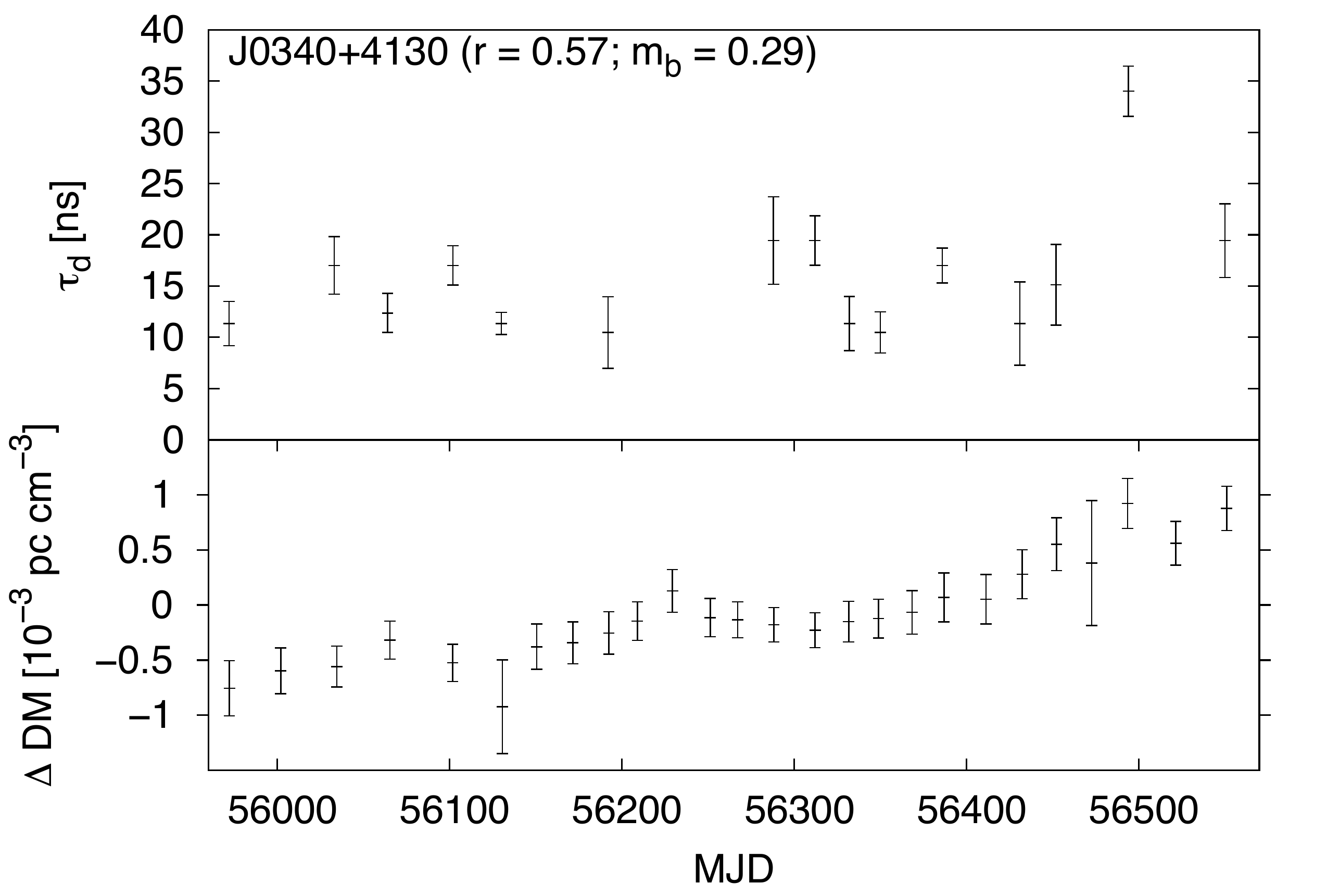}
		\includegraphics[width=0.32\textwidth]{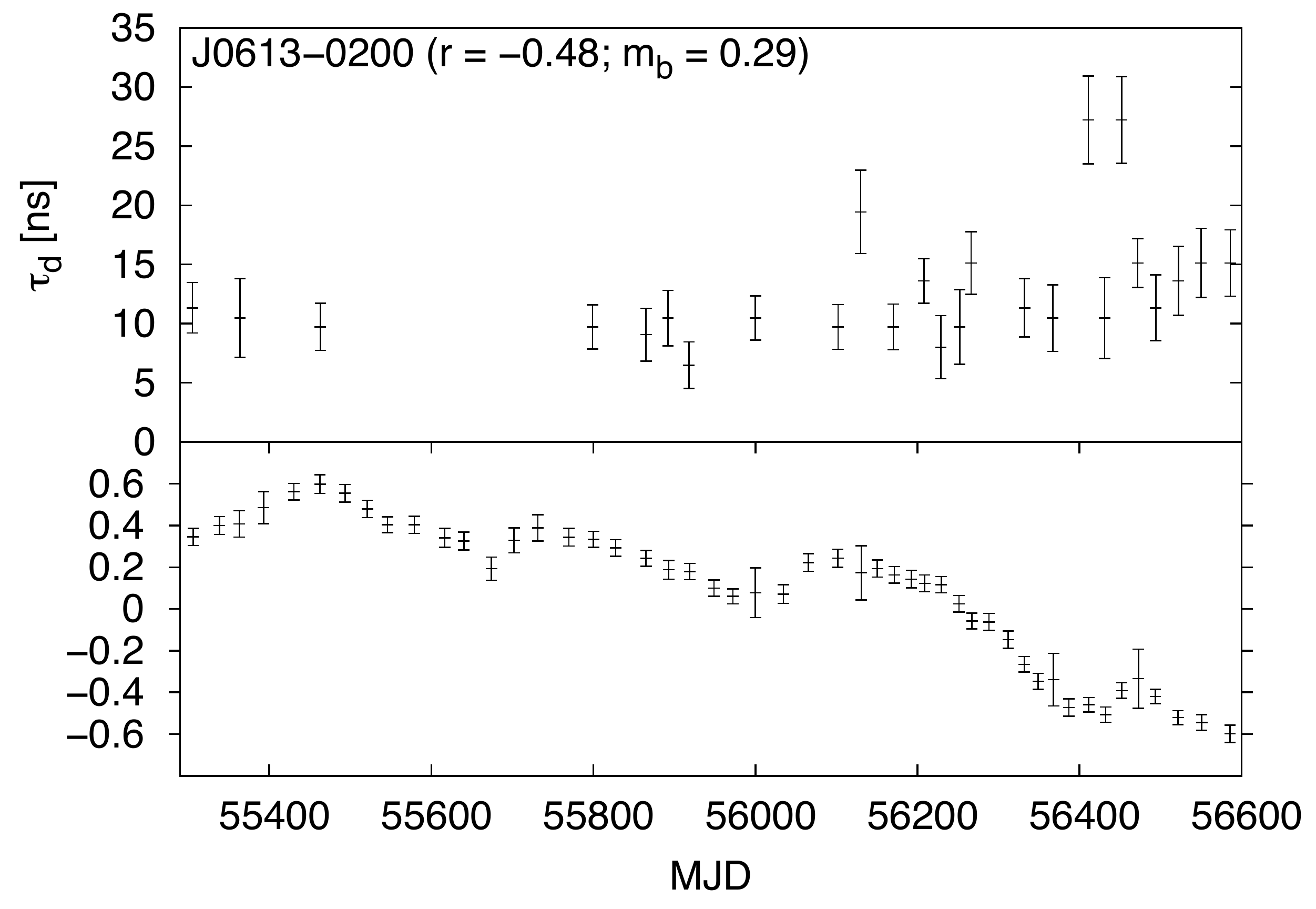}
		\includegraphics[width=0.32\textwidth]{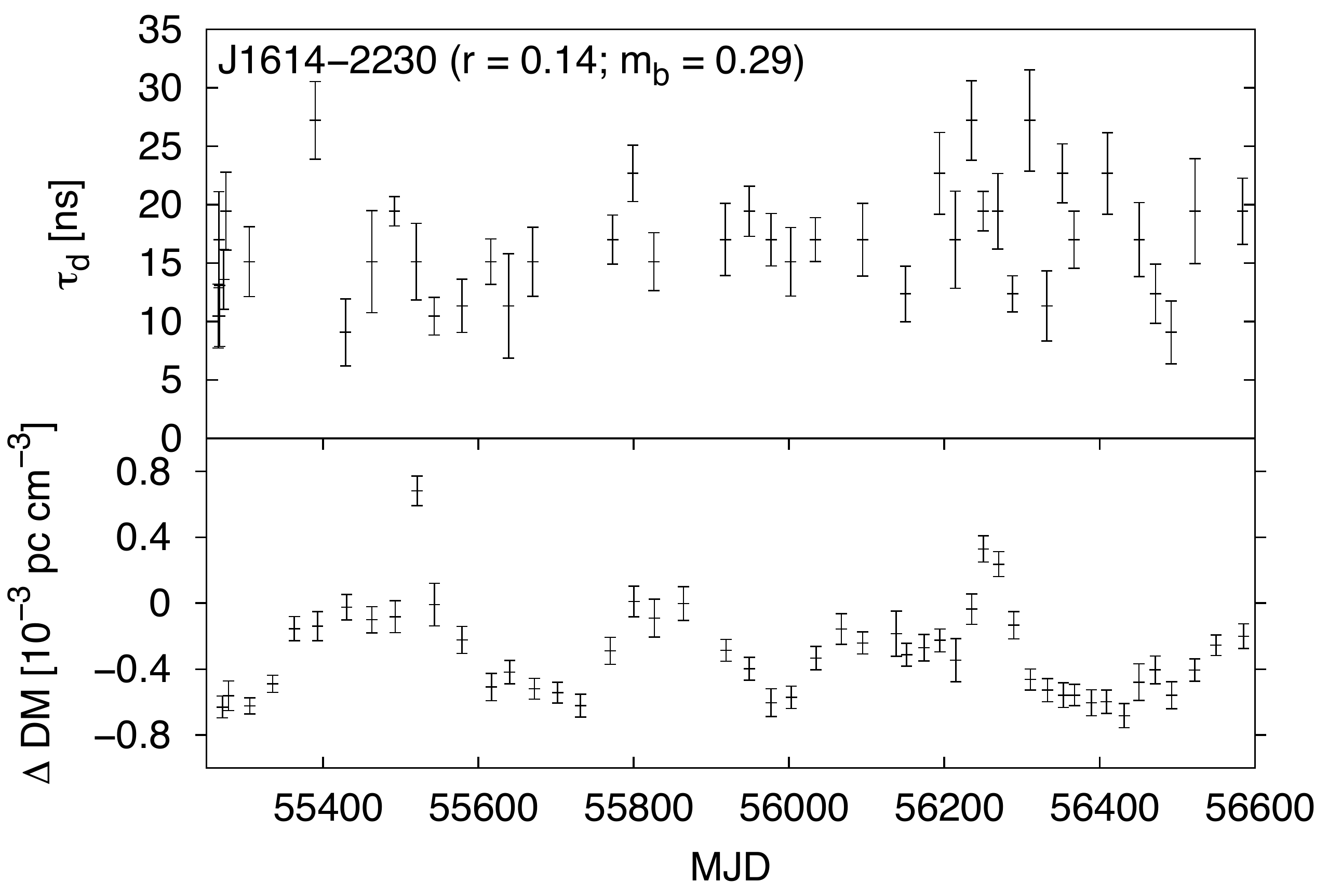}
		\includegraphics[width=0.32\textwidth]{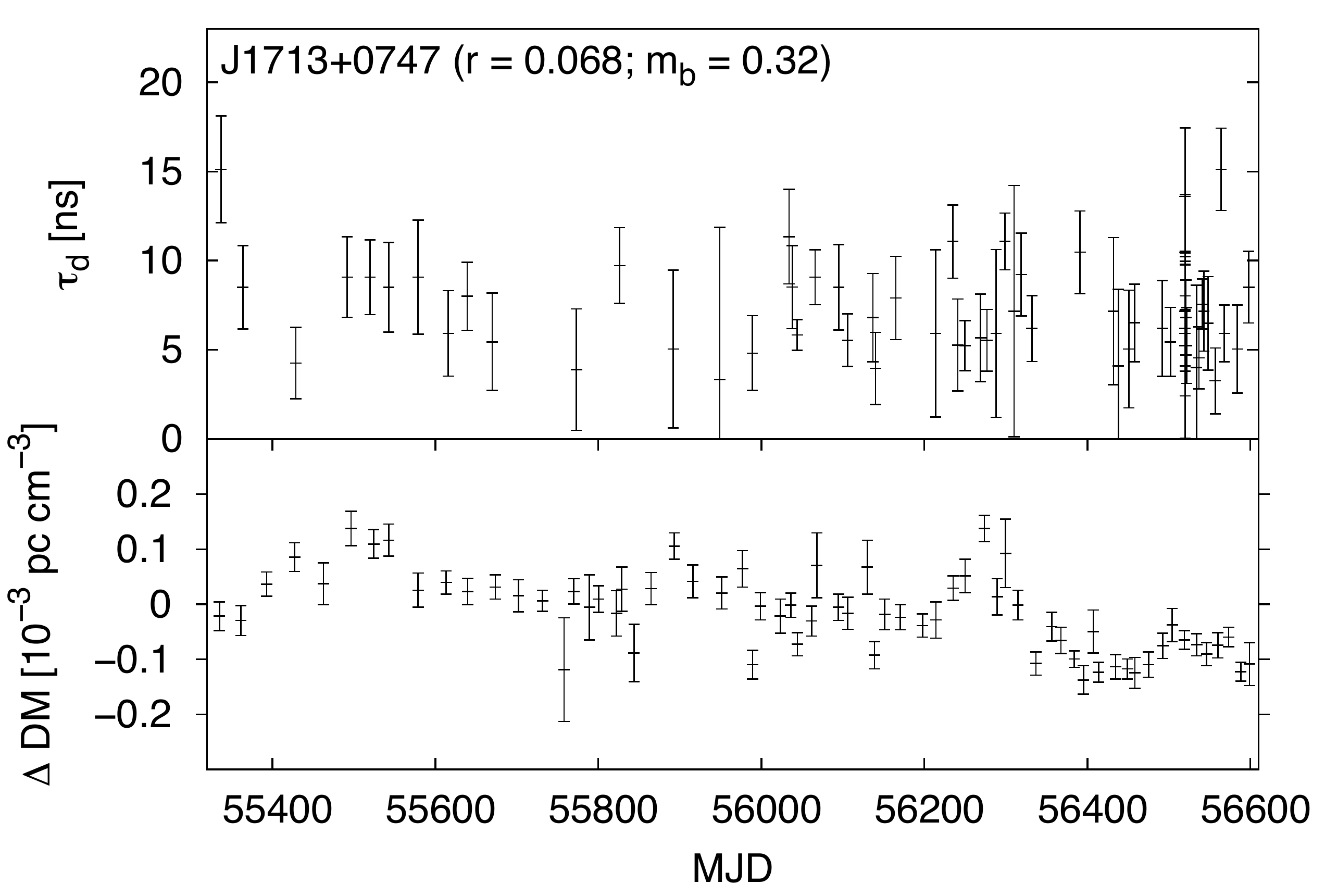}
		\includegraphics[width=0.32\textwidth]{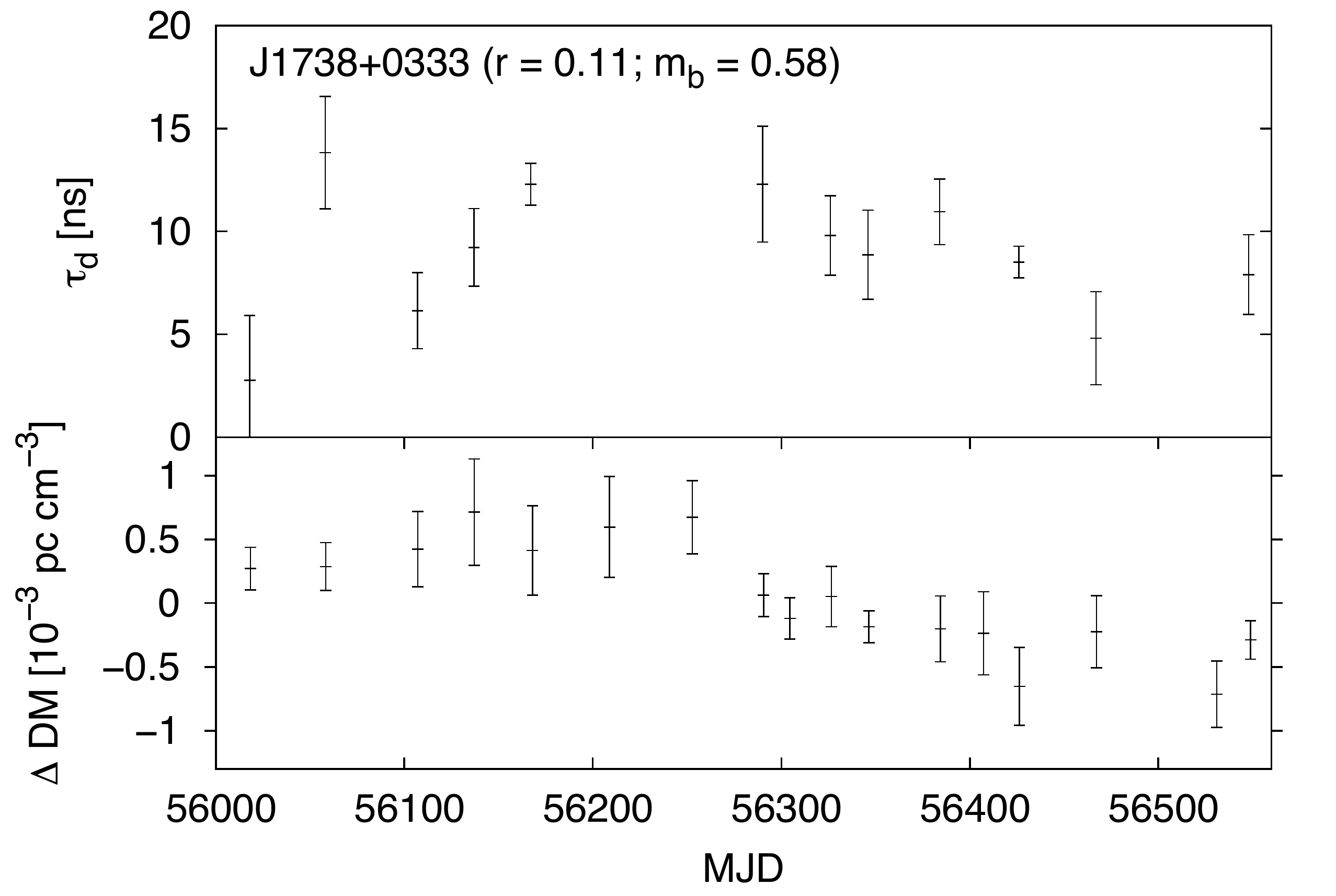}
		\includegraphics[width=0.32\textwidth]{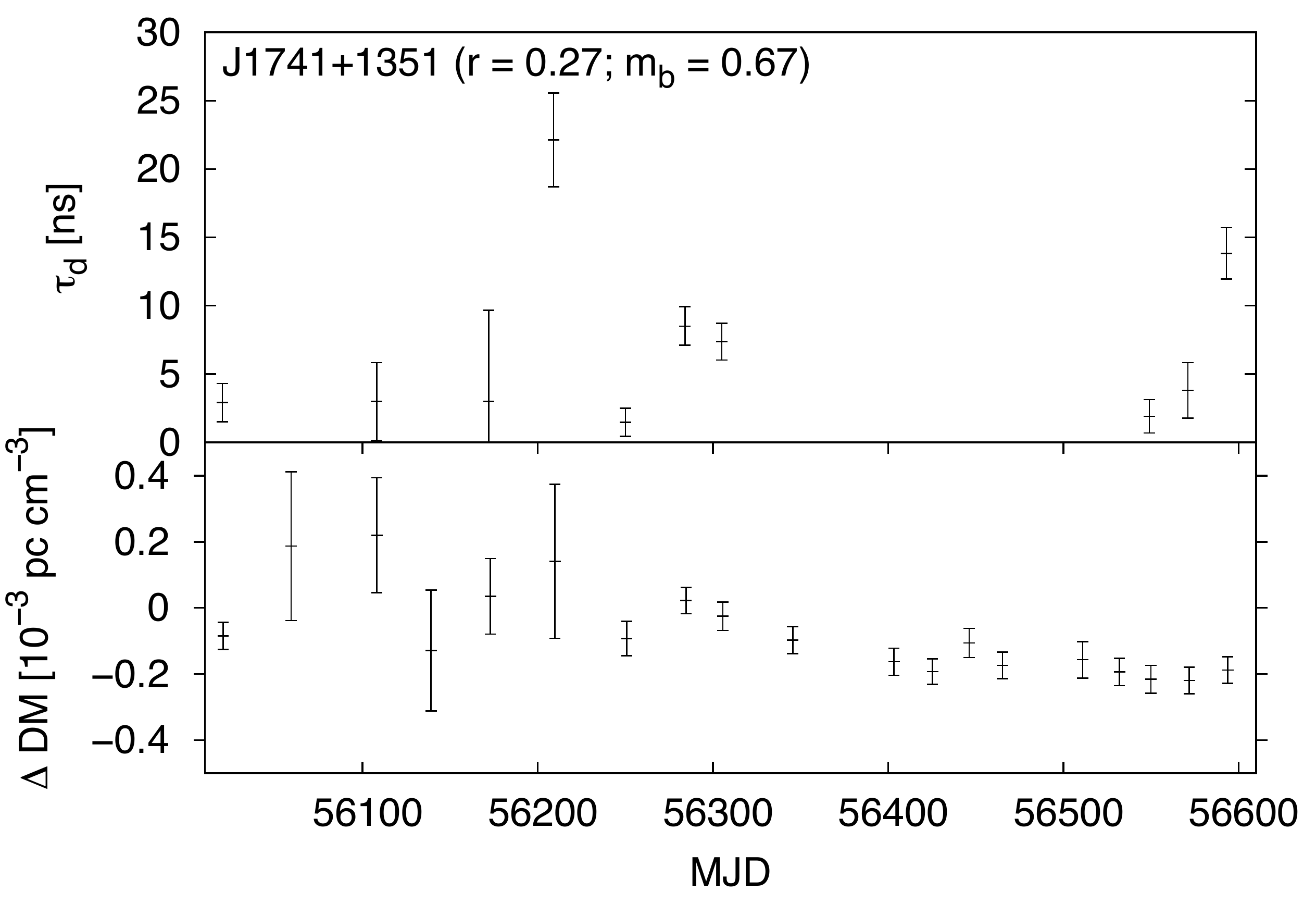}
		\includegraphics[width=0.32\textwidth]{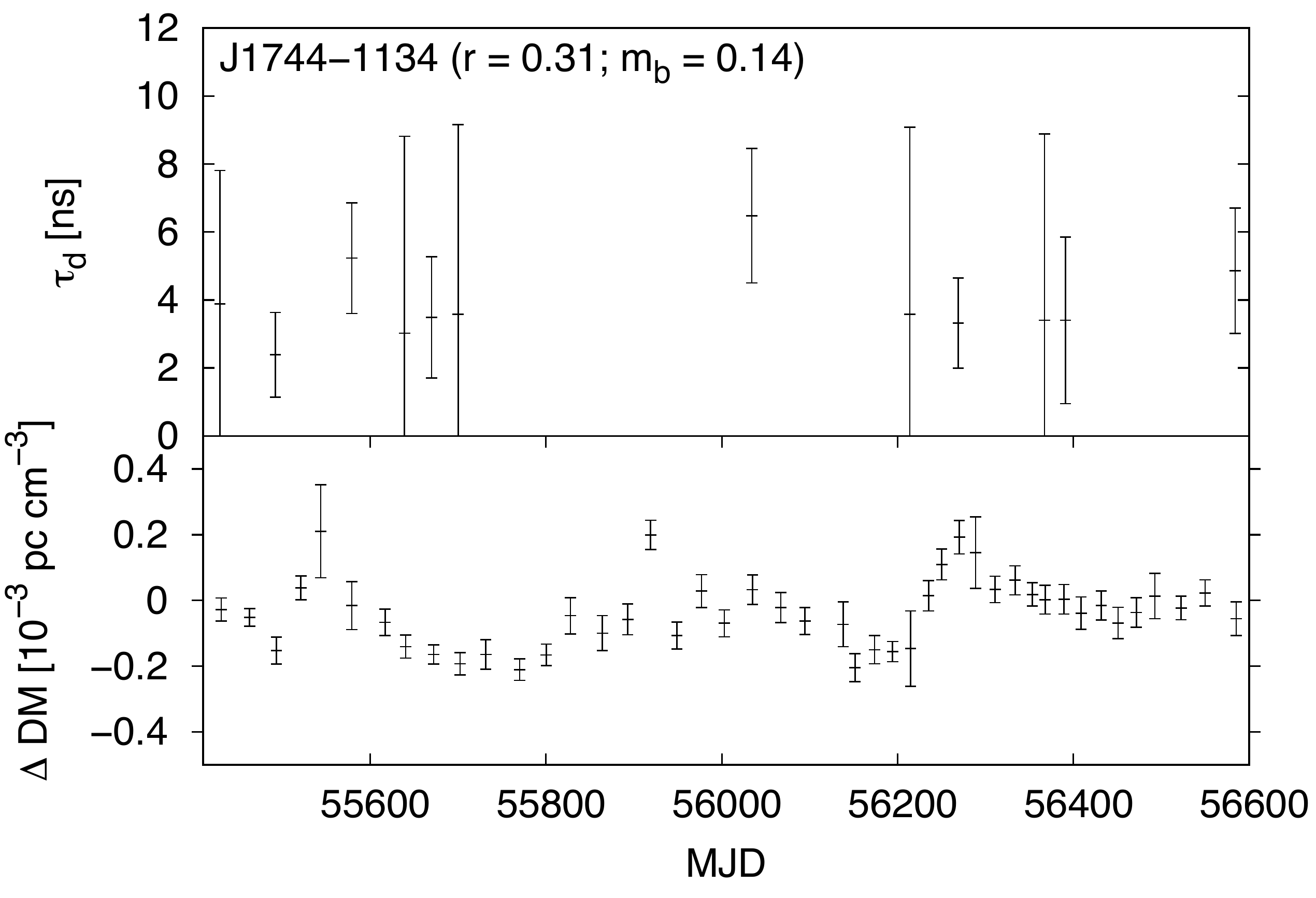}
		\includegraphics[width=0.32\textwidth]{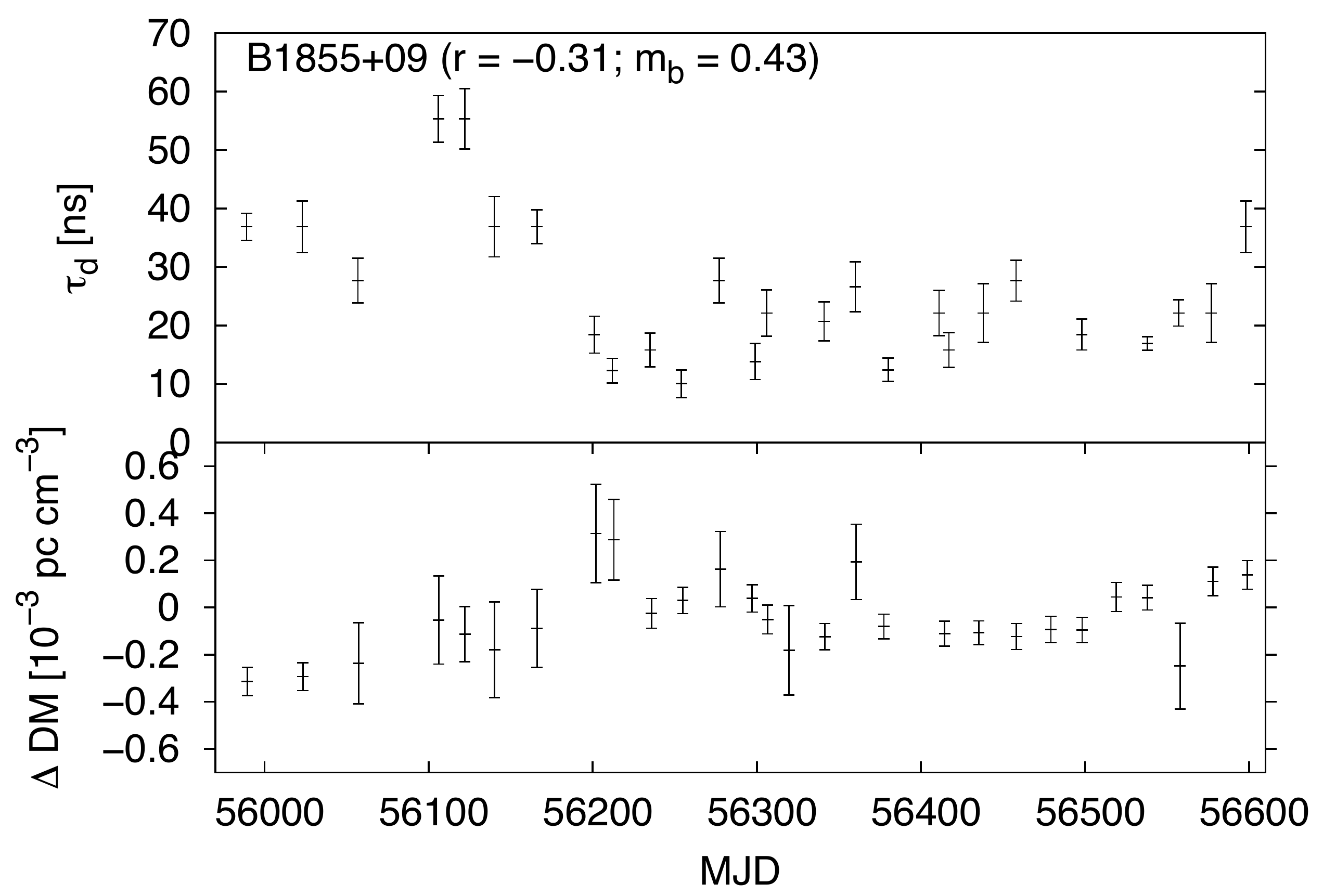}
		\includegraphics[width=0.32\textwidth]{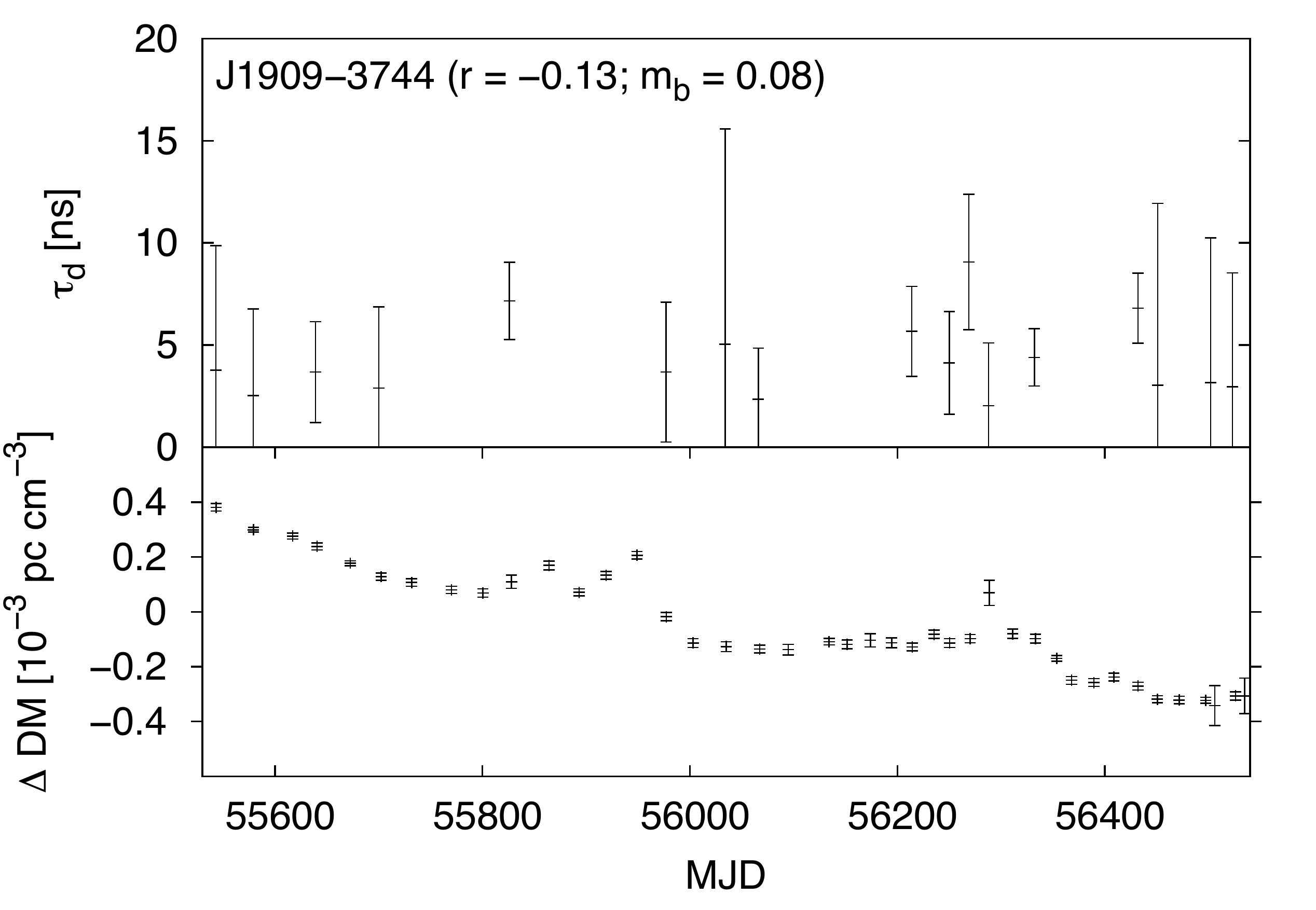}
		\includegraphics[width=0.32\textwidth]{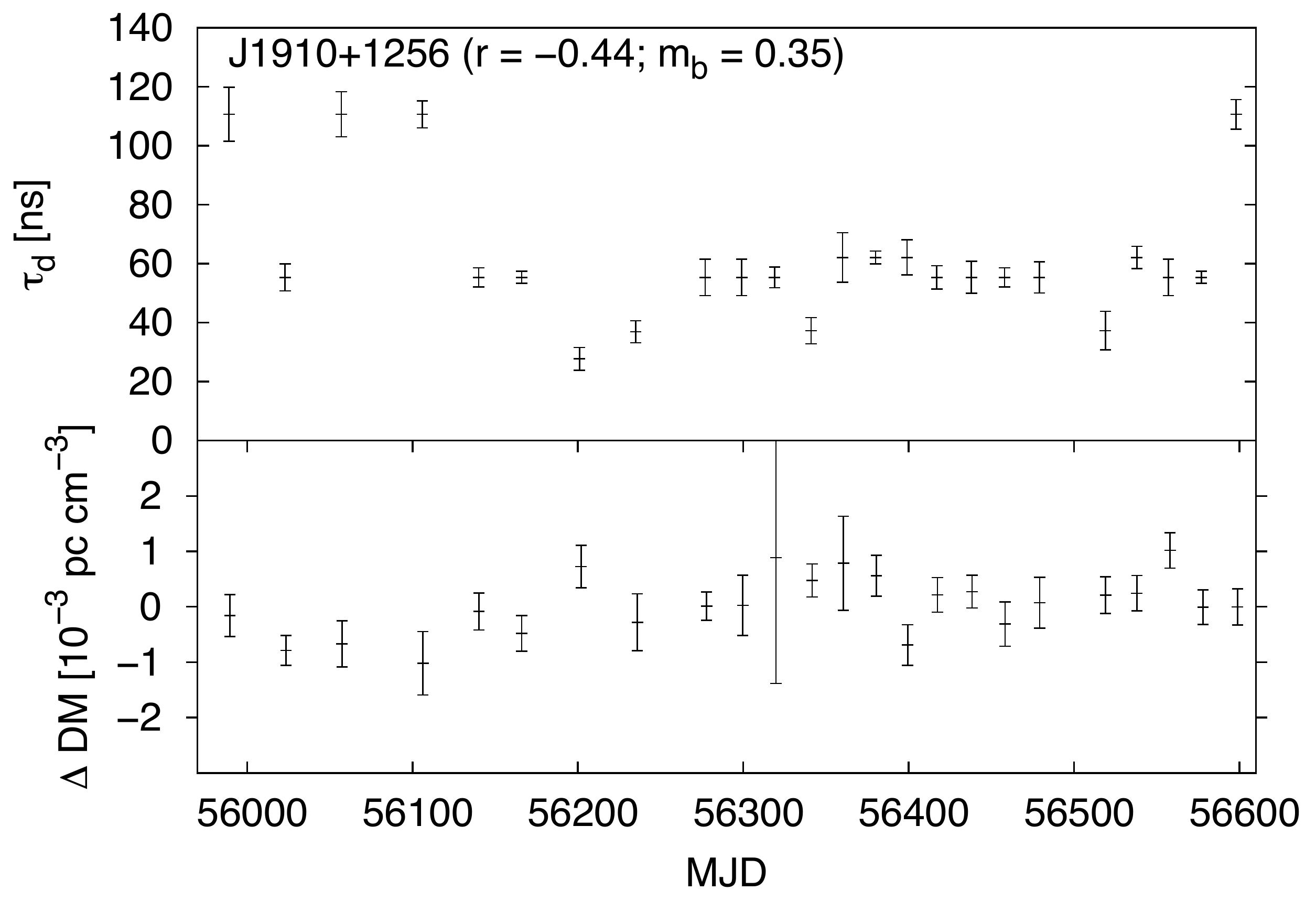}
		\includegraphics[width=0.32\textwidth]{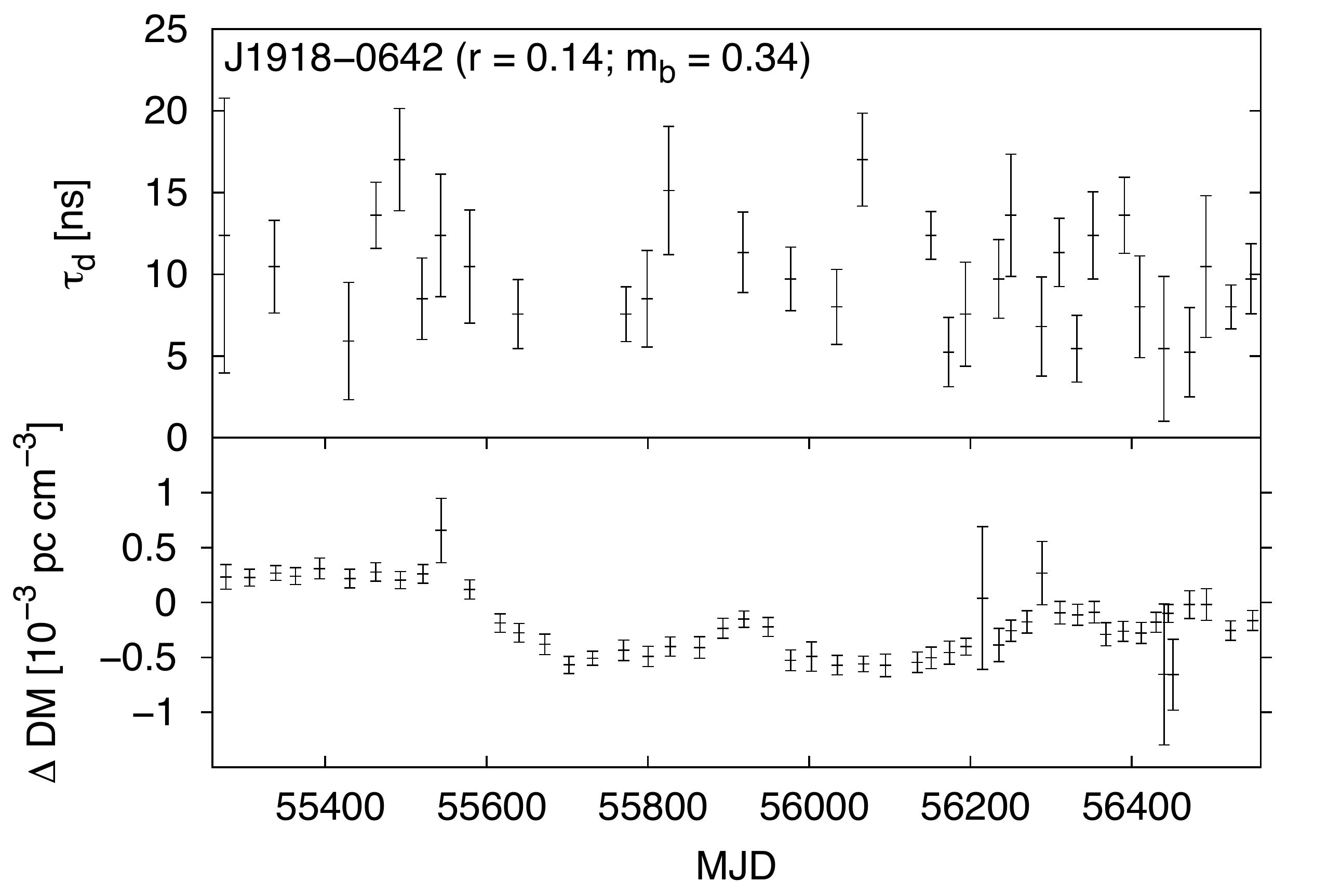}
		\includegraphics[width=0.32\textwidth]{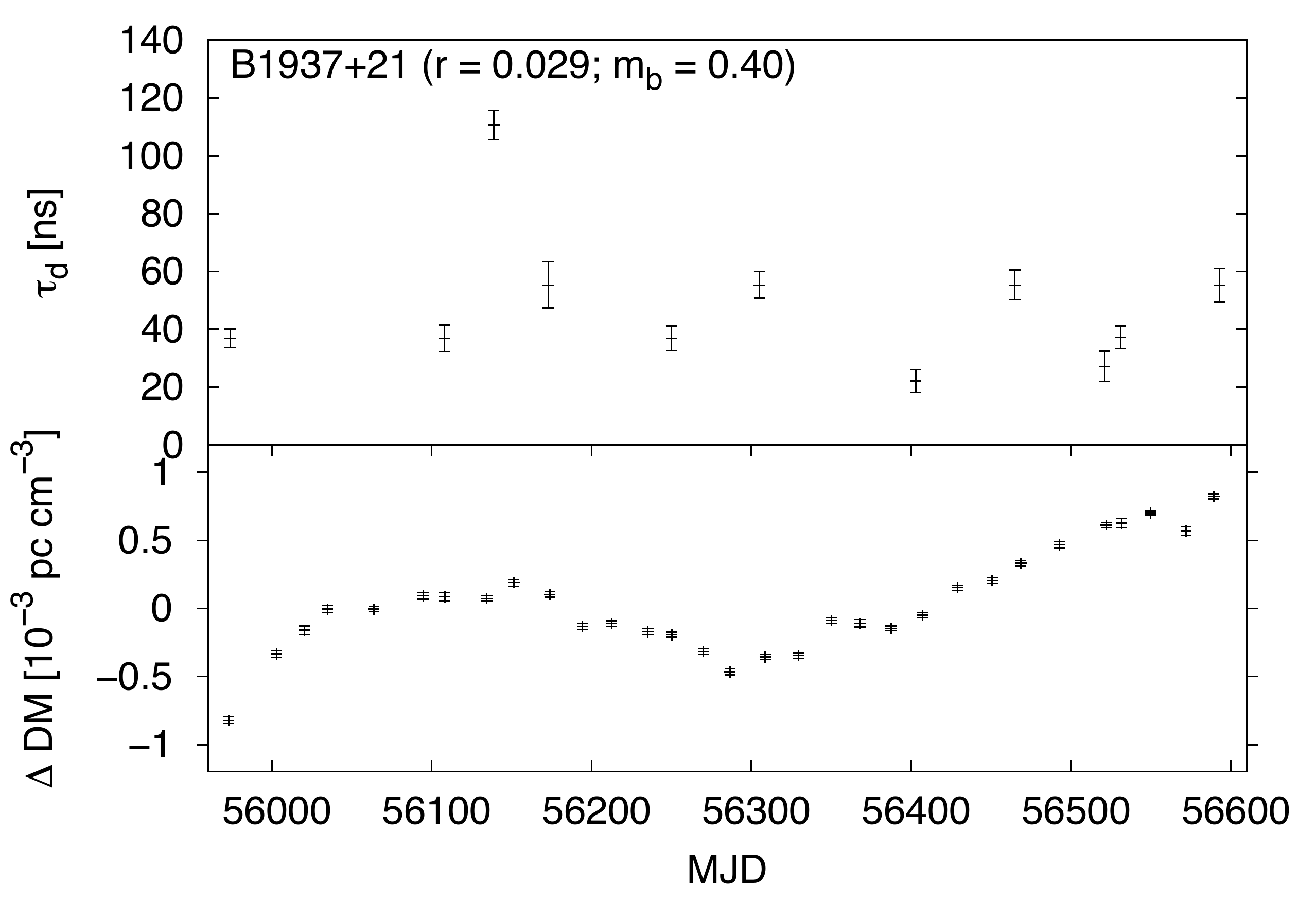}
		\includegraphics[width=0.32\textwidth]{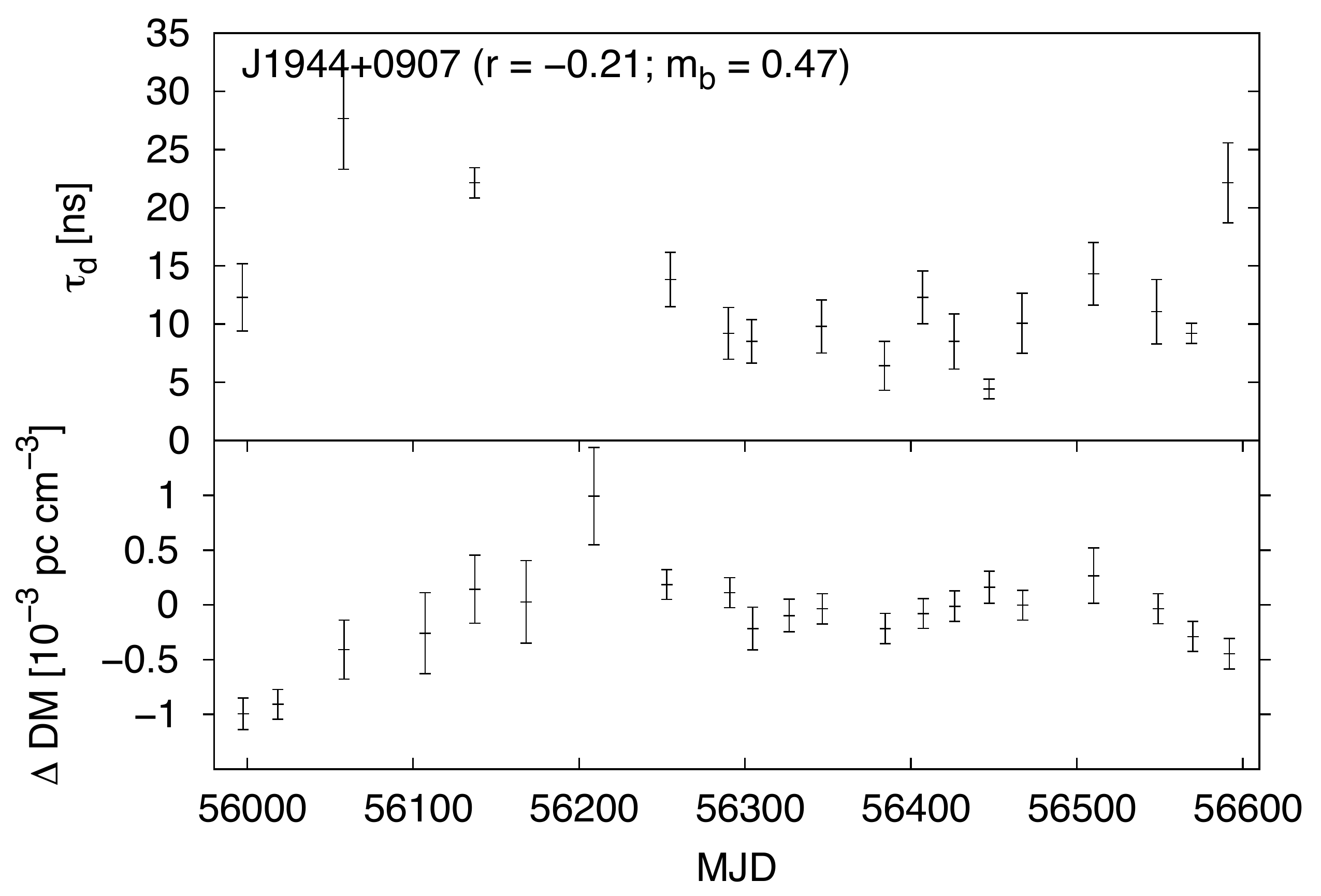}
		\includegraphics[width=0.32\textwidth]{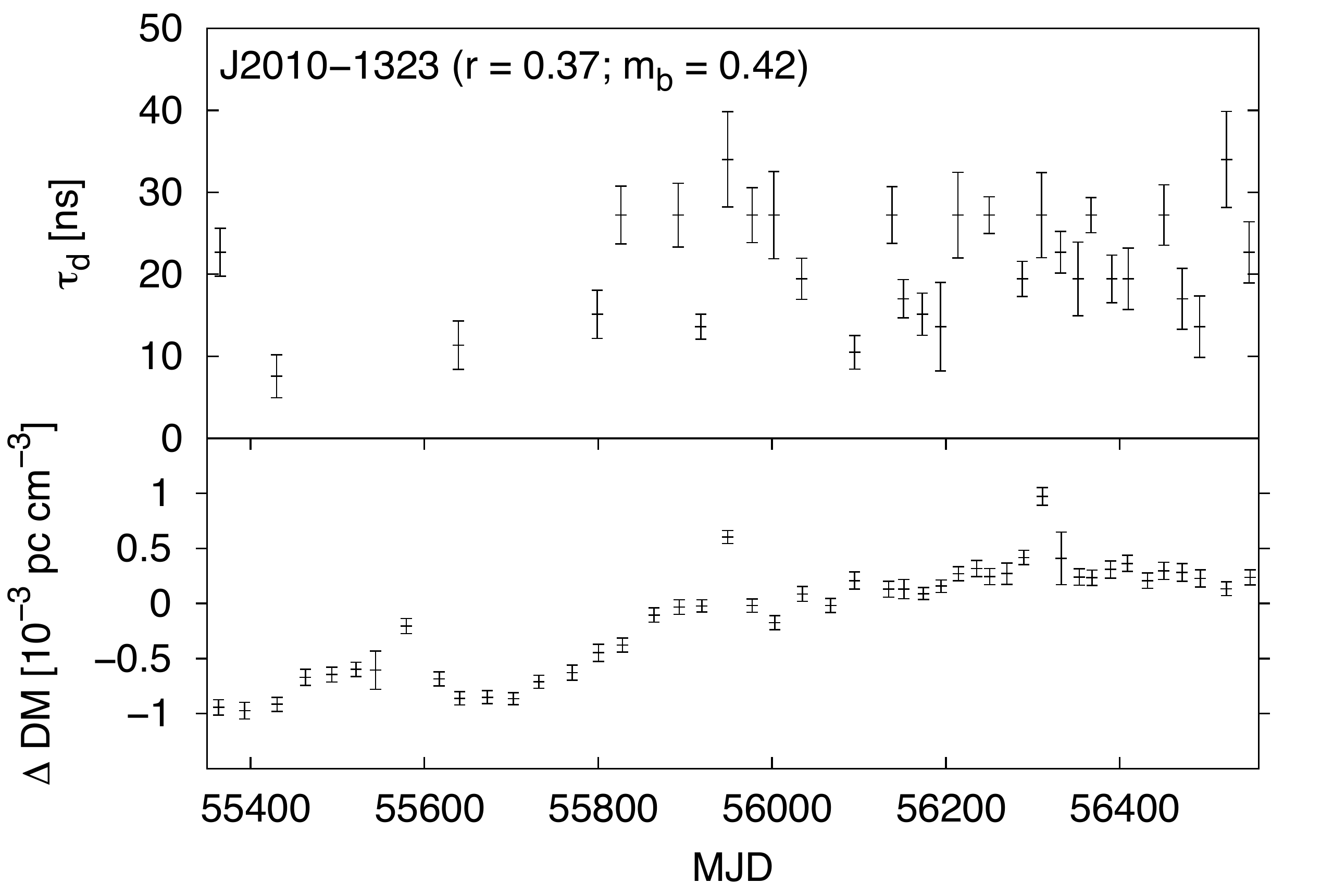}
		\includegraphics[width=0.32\textwidth]{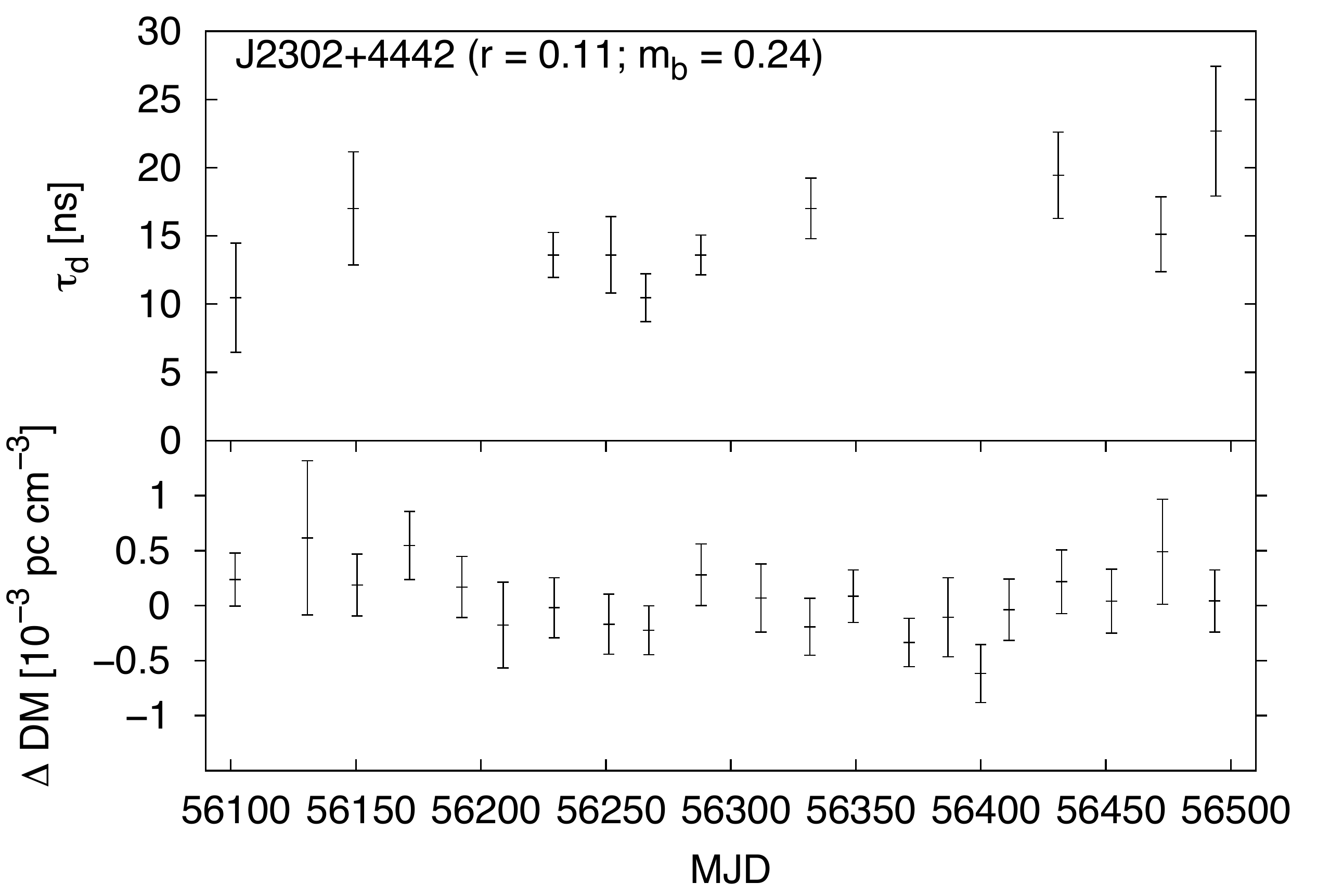}		
	\caption{Scattering delay variations (top panels) and DM variations (bottom panels) for pulsars with at least ten scattering delay measurements. All DM values measured within the same time span for each pulsar are included, even for epochs where no scattering delay measurement was possible. The DM values are measured in data from at least two separate frequency bands in the 9-year dataset. For each pulsar, the correlation coefficient, $r$, between the two variations is given in the parentheses in the top panel. In addition, the noise-corrected modulation index of the scintillation bandwidth is given as $m_{\rm b}$ for each pulsar (see Equations \ref{eq:mb} and \ref{eq:emb}).}
	\label{fig:delays}
	\end{center}
\end{figure*}

Two typical observations are shown in Fig.\,\ref{fig:dynspecex}, with the top dynamic spectrum displaying narrow scintles for PSR\,J1910+1256 with $\Delta \nu_{\rm d} \approx 3$\,MHz and the bottom dynamic spectrum showing wide scintles for PSR\,J2317+1439 with $\Delta \nu_{\rm d} \approx 34$\,MHz. Examples of scattering delay variability can be found in Fig.\,\ref{fig:delays}, where all NANOGrav pulsars with at least ten $\tau_{\rm d}$ measurements are plotted along with their DM variations. 

\section{Discussion}
\begin{table*}
  \begin{center}
  \caption{Previously published scintillation parameters.}  
  \label{tab:pub}
  \begin{tabular}{c |   l |   l |   l l l}
	\hline 
	\hline
	& {\bf This work} & {\bf NE2001} & \multicolumn{3}{l}{{\bf Previously published values}} \\
	\hline	
	Pulsar & $\bar{\tau}_{\rm d}$ & $\tau_{\rm d}^{\rm NE2001}$ & ${\tau}_{\rm d;1500MHz}$ & $\nu$ & Ref.\\
		   & (ns)          & (ns)         & (ns)             &  (MHz)    &   \\
	\hline
	J0030+0451 & --                     & 0.055 &  0.12 & 435 & \cite{nic01} \\ 
 	J0613--0200 & 11.7 $\pm$ 3.9 & 16 & 61.2 & 1369 & \cite{col10} \\
 	         $\shortparallel$       &             $\shortparallel$            &  $\shortparallel$ & 97.0$^{*}$ & 1500$^{*}$ & \cite{kei13} \\
	J1024--0719 &  2.8 $\pm$ 1.3 & 0.17 & 0.34 & 685 & \cite{col10} \\
 	         $\shortparallel$       &             $\shortparallel$            &  $\shortparallel$ & 0.59$^{*}$ & 1500$^{*}$ & \cite{kei13} \\
	J1455--3330 &  4.0 $\pm$ 1.1 & 0.94  & 0.51 & 436 & \cite{joh98} \\ 
	J1600--3053 &   -- 		     & 93      & 1768$^{*}$ & 1500$^{*}$ & \cite{kei13} \\
 	         $\shortparallel$       &             $\shortparallel$            &  $\shortparallel$ & 1072 & 3100 & \cite{col10} \\
	J1640+2224 & 2.6 $\pm$ 1.1 &   5.8   &  3.3 & 430 & \cite{bog02} \\
	J1643--1224 &   -- 		     & 90       & 7234$^{*}$ & 1500$^{*}$ & \cite{kei13} \\
 	         $\shortparallel$       &             $\shortparallel$            &  $\shortparallel$ & 2918 & 3100  & \cite{col10}\\
	J1713+0747 & 7.1 $\pm$ 2.4 & 4.1    & 1.1  & 430 & \cite{bog02} \\  
	         $\shortparallel$       &             $\shortparallel$            &  $\shortparallel$	 & 0.48  & 436  & \cite{joh98} \\  
	         $\shortparallel$        &            $\shortparallel$            &   $\shortparallel$       & 1.1    & 685  & \cite{col10} \\  
 	         $\shortparallel$       &             $\shortparallel$            &  $\shortparallel$ & 6.6 & 1500$^{*}$ & \cite{kei13} \\
	J1744--1144 & 3.8 $\pm$ 1.3 & 0.020 & 0.52  & 436 & \cite{joh98} \\ 
	          $\shortparallel$         &          $\shortparallel$              &     $\shortparallel$      & 1.87  &  660 & \cite{joh98} \\  
	          $\shortparallel$         &          $\shortparallel$             &     $\shortparallel$      & 0.40  & 685 & \cite{col10}\\ 
 	         $\shortparallel$       &             $\shortparallel$            &  $\shortparallel$ & 2.7$^{*}$ & 1500$^{*}$ & \cite{kei13} \\
	B1855+09 & 21.3 $\pm$ 9.9 & 4.0          &  6.5 & 430 & \cite{dew88} \\ 
	$\shortparallel$ &          $\shortparallel$                &    $\shortparallel$          & 1.1 & 685  & \cite{col10}\\
	$\shortparallel$ &          $\shortparallel$                &     $\shortparallel$          & 13.0 & 1369 & \cite{col10}\\
 	         $\shortparallel$       &             $\shortparallel$            &  $\shortparallel$ & 28.9$^{*}$ & 1500$^{*}$ & \cite{kei13} \\
	J1903+0327 & --                    & 2.3$\times 10^5$ & 9.3$\times 10^4$ & 1400 & \cite{cha08} \\
	J1909--3744 & 4.9 $\pm$ 1.8 & 1.5      & 0.68 & 685  & \cite{col10}\\
 	         $\shortparallel$       &             $\shortparallel$            &  $\shortparallel$ & 4.3$^{*}$ & 1500$^{*}$ & \cite{kei13} \\
	B1937+21 & 44.3 $\pm$ 21.4 & 130       & 127 & 320 & \cite{cor90} \\
	$\shortparallel$ & $\shortparallel$  &   $\shortparallel$      & 155 & 430 & \cite{cor90} \\ 
	$\shortparallel$ & $\shortparallel$  &   $\shortparallel$      & 48.4 & 1369 & \cite{col10}\\ 
	$\shortparallel$ & $\shortparallel$  &   $\shortparallel$      & 127 & 1400 & \cite{cor90} \\ 
 	         $\shortparallel$       &             $\shortparallel$            &  $\shortparallel$ & 132$^{*}$ & 1500$^{*}$ & \cite{kei13} \\
	J2145-0750 & 2.8 $\pm$ 0.7 & 0.53   & 0.47 & 436 & \cite{joh98} \\
	$\shortparallel$ &         $\shortparallel$             &   $\shortparallel$        & 0.45 & 685 & \cite{col10}\\
 	         $\shortparallel$       &             $\shortparallel$            &  $\shortparallel$ & 0.82$^{*}$ & 1500$^{*}$ & \cite{kei13} \\
	J2317+1439 & 3.0 $\pm$ 1.0 & 1.9     & 1.4 & 436 & \cite{joh98} \\
	\hline
	\hline
	\multicolumn{6}{l}{{\bf Notes.} The published values are reported at the given observing frequency ($\nu$) and here} \\
	\multicolumn{6}{l}{converted to scattering delays (${\tau}_{\rm d;1500MHz}$) at 1500 MHz using a scaling index of $\zeta = 4.4$.} \\
	\multicolumn{6}{l}{For reference, scattering delay values from this work is included as $\bar{\tau}_{\rm d}$ and scaled values}\\
	\multicolumn{6}{l}{from the NE2001 are given as $\tau_{\rm d}^{\rm NE2001}$.}\\
	\multicolumn{6}{l}{$^{*}$Only an already scaled value is reported in the original publication.}
  \end{tabular}
  \end{center}
\end{table*}

Previous work on scintillation properties of pulsars has been carried out at various radio frequencies. 
To compare the scattering delays in this paper with previously published values, we need to scale all values to a common observing frequency. 
This introduces large uncertainties, since the comparison will be largely dependent on the chosen scaling law and, as shown in this paper, the measured scaling indices are not only different for different pulsars and hence observing directions, but may also vary with time (as discussed in more detail in Sec \ref{sec:scalingindex}).
Nevertheless, a comparison with previously published values is listed in Table \ref{tab:pub}, by scaling the published scattering delays to their value at an observing frequency of 1500\,MHz, using a scaling index $\zeta = 4.4$. 

In some cases, the previously published values show large discrepancies with the values in this work. 
\cite{col10} and \cite{kei13} both list scintillation parameters for millisecond pulsars in the Parkes Pulsar Timing Array, some of which are also included in the NANOGrav array. 
The values in \citeauthor{kei13} have already been scaled to a common observing frequency of 1500\,MHz, while \citeauthor{col10} list the actual frequency of each observation. 
Some of the values from \citeauthor{col10} were measured at an observing frequency $\nu_{\rm obs}$ = 1369\,MHz, and these values agree better with our measurements than the rest of the pulsars in \citeauthor{col10}, which suggests that the choice of scaling index may be an issue. 
Another explanation for the discrepancy is the variation of scattering delay over time (see Fig.\,\ref{fig:delays}), which is also observed by \citeauthor{col10} 
Similarly, \cite{bha98} observed variations of the scintillation bandwidth of a factor of 3$-$5 when analyzing scintillation parameters of slow low-DM pulsars.
In the data presented here, we observe delays with a variation of up to an order of magnitude.

In the cases where the published values are $\tau_{\rm d}$\,$\lesssim$\,1\,ns, the values measured in this work are always larger. This is likely due to the limited bandwidth of the data, which allows only the observations with higher scattering delay values in the distribution to be resolved.
Many of the values published before 2001 are very similar to the corresponding value from the NE2001 model. This is expected, since in the creation of the electron density model, as many scattering delay measurements as possible were included \citep{cor02} and hence these early measurements were likely used as input to the model.

\subsection{Comparison with the NE2001 model}
The most comprehensive and frequently used model of the free electrons in the Galaxy is the NE2001 model \citep{cor02}. In addition to predicting distances to pulsars from their DM values, this model also estimates the scintillation properties from a position and a DM value. The scattering delays predicted for the NANOGrav pulsars by the model are listed in Table \ref{tab:allscatt}. 

\begin{figure}
	\begin{center}
		\includegraphics[width=0.45\textwidth]{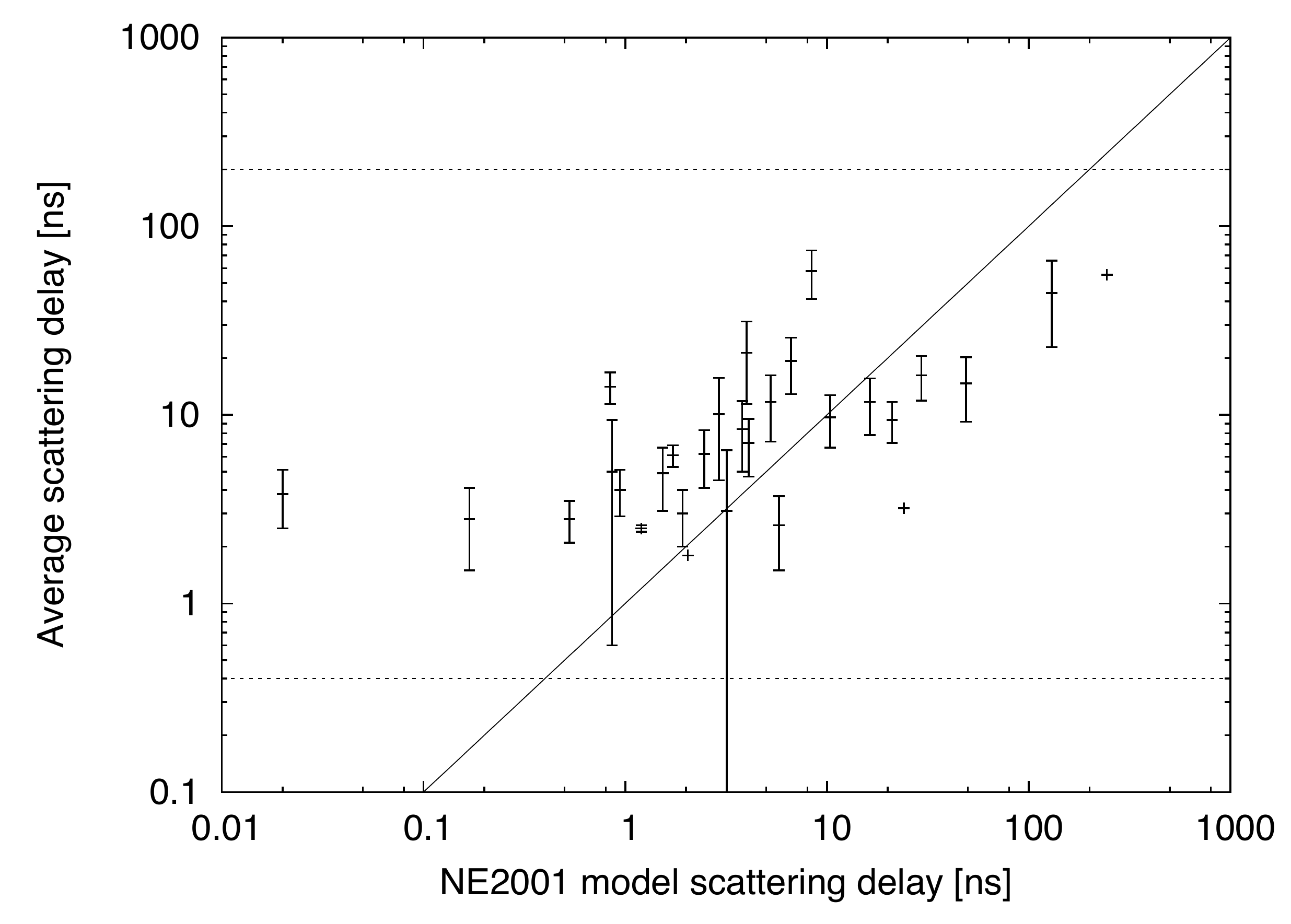}
	\caption{Comparison of average scattering delay to the estimated NE2001 model values. The NE2001 values have been scaled to an observing frequency of 1500 MHz, using a scaling index $\zeta = 4.4$. The two horizontal dotted lines mark maximum and minimum scattering delays that can be measured from the data. These limits are calculated from the frequency resolution and the total bandwidth of the data respectively, using the definition of the scintillation bandwidth as the half-width at half-maximum of the Gaussian fit to the ACF. The diagonal line represents equality of model and measurement.}
	\label{fig:NE2001}
	\end{center}
\end{figure}

These values have also been plotted against the average measured scattering delays in Fig \ref{fig:NE2001}. The horizontal dotted lines show the minimum and maximum measurable delays in the data, while the diagonal line shows equality of the measurement to the model. In general, the measured values are slightly higher than the predictions. It is known that the model predictions have large associated uncertainties, in particular at higher Galactic latitudes, where the numbers of sources with known distances and DM are low. 

For some of the pulsars, no measurement was possible, e.g., due to too narrow or too wide scintles. All but two of these non-detections are predicted to lie close to the detection limits of the data, and hence it is not surprising that no scattering delays have been measured for these pulsars. One exception is PSR\,J0645+5158, which has $\tau_{\rm d;NE2001}  = 6.2$\,ns at 1500\,MHz.  
Manual inspection of the dynamic spectra obtained for this pulsar shows that in the few observations with scintles strong enough to be detected, radio interference contamination caused errors in the ACF calculation.
By eye, we estimate a scintillation bandwidth of $\nu_{\rm d} \approx 100$\,MHz, which corresponds to $\tau_{d} \approx 1.5$\,ns.
The other exception is PSR\,J1832--0836, with $\tau_{\rm d;NE2001}  = 18$\,ns at 1500\,MHz. This pulsar is a recent addition to the NANOGrav array, and only 10 observations at 1500\,MHz are included in the 9-year data release. 
Even by manually inspecting the dynamic spectra for PSR\,J1832$-$0836, no scintles are detected, which is likely due to the low flux density of the signal.

For the pulsars with $\tau_{\rm d; NE2001} \ll 1$\,ns, the measured scattering delay is much larger than the model delay. 
It is possible that there is an underlying structure of wide scintles that are too wide to detect in this dataset, and what is measured here are small modulations on top of the larger structure. Another possibility is that the NE2001 model largely underestimates the amount of scattering in the direction of these pulsars, which are all at high Galactic latitudes where the model is known to have large uncertainties \citep[e.g.][]{gae08, cha09}. 
For the cases where the $\tau_{\rm d; NE2001}$-values are close to 1\,ns, we may only be detecting the highest values in the distribution of delays. Since the bandwidth of the observations is limited, the true mean $\tau_{\rm d}$-values may be lower than what we can measure. 
A similar argument can be applied to the $\tau_{\rm d; NE2001}$-values close to the maximum delay limit, where the limited frequency resolution would permit detection of only the lower parts of the delay distributions. In addition, the delays in particular observations may be underestimated for pulsars close to the maximum limit, since a strong scintle with true $\Delta \nu_{\rm d}$ smaller than the channel bandwidth would still be measured as being one channel wide.

\subsection{Interstellar medium variations}
The measured values of $\tau_{\rm d}$ are plotted in Fig.\,\ref{fig:delays} as a function of observing day for all pulsars with at least 10 measured scattering delay values. 
To analyze the delay variability we have calculated a modulation index for the scintillation bandwidth for each pulsar, defined as the RMS fluctuations of $\Delta \nu_{\rm d}$ divided by its average value \citep{bha99} and given by
\begin{equation}
	m_{\rm b; measured} = \frac{1} {\langle \Delta \nu_{\rm d} \rangle} \left( \frac{1}{N_{\rm obs} - 1}  \sum_{i=1}^{N_{\rm obs}}{(\Delta \nu_{\rm d, i} - \langle \Delta \nu_{\rm d} \rangle) ^2} \right) ^{1/2}
	\label{eq:mb}
\end{equation}
where $\langle \Delta \nu_{\rm d} \rangle$ is the average scintillation bandwidth, $N_{\rm obs}$ is the number of measurements, and $\Delta \nu_{\rm d, i}$ is the scintillation bandwidth at the $i$th epoch. 
The measurement error on $\Delta \nu$ will induce an apparent increase in the modulation indices, that needs to be corrected for. Following \cite{bha99}, we calculate corrected modulation indices as 
\begin{equation}
	m^2_{\rm b; corrected} = m^2_{\rm b; measured} - m^2_{\rm b; noise}
	\label{eq:emb}
\end{equation}
where the error induced modulation index, $m_{\rm b; noise}$, is calculated as the typical uncertainty in $\Delta \nu_{\rm d}$. 
The corrected values of $m_{\rm b}$ are given in the upper panel for each pulsar in Fig.\,\ref{fig:delays}. For two of the pulsars (J1744--1124 and J1909--3744), $m_{\rm b; noise} \approx m_{\rm b; measured}$. Both of these pulsars have only a small numbers of scintles in each observation, which contributes to the large uncertainty in the scattering delay measurements, as seen in Fig.\,\ref{fig:delays}, and hence are excluded from further modulation index analyses. For the remaining pulsars, $m_{\rm b; noise} \ll m_{\rm b; measured}$, so the corrected modulation is approximately equal to the measured modulation.

These modulations of the scintillation bandwidths, and hence the scattering delays, could originate from refractive scintillation effects. 
The theoretical model of the observable modulation effects due to refractive scintillation are highly dependent on the form of the density spectrum, which can be written as 
\begin{equation}
	P_{\delta n_{\rm e}} (\kappa) = C_{\rm n}^2 \kappa ^{-\beta}
\end{equation}
for values of the spatial wavenumber, $\kappa$, between inner and outer cutoffs in spatial scale sizes of the density irregularities in the ISM  \citep{arm95}. Here $C_{\rm n}^2$ represents the scattering strength and $\beta$ is the density spectral index, which for a Kolmogorov spectrum is equal to 11/3.  For $\beta < 4$, the corresponding scaling over the observed frequency band is
\begin{equation}
	\zeta = \frac{2\beta}{\beta-2}
\end{equation}
and hence for a Komogorov medium, $\zeta = 4.4$ \citep{cor10}. 
The scattering measure is defined as 
\begin{equation}
	{\rm SM} = \int^D_0{C^2_n ds}
\end{equation}
which can be simplified to 
\begin{equation}
	{\rm SM} = 292 \left( \frac{ \tau_{\rm d} }{D} \right) ^{5/6} \nu_{\rm GHz}^{11/3}
	\label{eq:SM}
\end{equation}
with units kpc m$^{-20/3}$ if the scattering delay, $\tau_{\rm d}$, is given in seconds and the distance, $D$, is given in kpc \citep{cor91}.

\cite{rom86} presented a theory for the refractive effects in a few special cases using power-law forms of density spectra with different spectral indices. In the case of a Kolmogorov spectrum and considering scattering in a thin screen halfway between the observer and the pulsar, the modulation index for the scintillation bandwidth is given by
\begin{equation}
	m_{\rm b; Kolmogorov} \approx 0.202 (C^2_{\rm n})^{-1/5} \nu_{\rm obs}^{3/5} D^{-2/5}
\end{equation}
where $\nu_{\rm obs}$ is the observing frequency in GHz, $D$ is the distance to the pulsar in kpc and $C^2_{\rm n}$ is the scattering strength in units of $10^{-4}$m$^{-20/3}$ \citep{bha99}. For a Kolmogorov spectrum,
\begin{equation}
	C^2_{\rm n} = 0.002 \Delta \nu_{\rm d}^{-5/6} D^{-11/6} \nu_{\rm obs}^{11/3} {\rm m}^{-20/3}
\end{equation}
where $\Delta \nu_{\rm d}$ is given in MHz, $D$ is given in kpc and the observing frequency, $\nu_{\rm obs}$, is given in GHz \citep{ric77,cor86b}. 
Calculating the theoretical modulation indices for the pulsars in Fig.\,\ref{fig:delays}, 
using pulsar distances from parallax measurements where available\footnote{With values as given in the ATNF pulsar catalogue: http://www.atnf.csiro.au/research/pulsar/psrcat/} and otherwise as given by the NE2001 model \citep{cor02}, results in values ranging $0.15 \le m_{\rm b} \le 0.20$. 
If the density spectrum is not Kolmogorov in shape, but rather has $\beta > 11/3$, the fluctuations are expected to be independent of the observing wavelength, but give larger modulation than a Kolmogorov spectrum. For $\beta = 4.0$ the theoretical $m_{\rm b} \approx 0.35$ and for $\beta = 4.3$ the theoretical $m_{\rm b} \approx 0.57$ \citep{rom86, bha99}. 

Comparing the corrected modulation indices with the theoretical ones, all of the measured $m_{\rm b}$-values are larger than those predicted for a Kolmogorov medium. This does however not necessarily mean that the density spectrum is not Kolmogorov. Instead, if the ISM in these directions is not well described by a thin screen, but rather as an extended Kolmogorov medium, that would explain the larger observed modulation. 

We can also approach this discussion from the other direction, by looking at the measured scattering delay frequency scaling. Many of the measured scaling indices in Table \ref{tab:scalingindex} are smaller than that expected for a Kolmogorov medium, even within the errors, and similar trends have been shown in previous work by other authors.  
In an effort to measure pulse scattering broadening of slow pulsars, \cite{bha04} observed a number of pulsars at a minimum of two different frequencies with the Arecibo telescope. Indeed, for most of their sources, they measured a scaling index $\zeta <$ 4.4. By using their results together with previously published values, \citeauthor{bha04} inferred an empirical frequency scaling index of 3.9$\pm$0.2. Following the discussion above, the smaller observed scaling indices agree well with the larger modulation observed in our data.

\subsection{Dispersion Measure variations}
The only ISM effect usually corrected for in high-precision pulsar timing is the DM and its variations. Since dispersion and scintillation both arise from the same interstellar structures in each observation, it is tempting to assume a correlation between the variations of the two delays. We have compared the scattering delay and DM variations by calculating correlation coefficients for all pulsars in Fig.\,\ref{fig:delays}. 
The DM values are measured using the DMX method in the pulsar timing software {\sc tempo}\footnote{http://tempo.sourceforge.net}. In this procedure, the DM(t) function is treated as piecewise constant, and an independent DM value is fitted for over successive 14-day windows, in a simultaneous fit with all other timing model parameters. 
This is similar to the procedure used in \cite{arz15}, except here only data from the GUPPI and PUPPI backends are used. 
It is known that timing using the DMX method absorbs some of the other ISM effects in addition to the DM variation (see Section \ref{sec:timing} and \cite{dem13}). Even so, some correlation between the two variations could be expected.
A correlation coefficient, $r$, is given for each pulsar in their respective panel of Fig.\,\ref{fig:delays}, calculated by comparing DM values and scattering delays from epochs with a measured scattering delay only. 
None of the pulsars show any convincing correlation. Some of them even show trends of being anti-correlated, which is probably a result of small number statistics. 
The highest possibilities of correlation between the DM and scattering variations are for pulsars with fewer measurements, suggesting that the $r$-value might decrease with the addition of more data points. 

Recently, \cite{col15} found so-called Extreme Scattering Events \citep[ESEs; ][]{fie87} in data collected for the Parkes Pulsar Timing Array.  These are characterized by a sudden, large change in DM occurring simultaneously with a decrease in scintillation bandwidth. We find no such events in the data presented here. 
A more comprehensive study of DM variations of the NANOGrav MSPs will be published elsewhere (Jones et al., 2015, {\it in prep}).

\subsection{Modulations of the frequency scaling index}
\label{sec:scalingindex}
For some of the pulsars in the sample, we have measured the frequency scaling at multiple epochs. The resulting values are listed in Table \ref{tab:scalingindex}, together with the MJDs of the relevant observations. A large range of scaling values were measured, varying between observing epochs for the same pulsar. 
To investigate the variation further, we chose to focus on the measurements for PSR\,J1614--2230, for which the resulting scaling values vary between $2 \lesssim \zeta \lesssim 6$. 
When measuring the scaling, the entire observing band is divided up in four sub-bands, as described in Sec.\,\ref{sec:analysis}. Generally, only a small number of scintles are found in each of the sub-bands, implying large uncertainties on the individual measurements of $\Delta \nu_{\rm d}$, which could be an issue when estimating the scaling. 
Another complication with a small number of scintles is that the result might be biased by where within each sub-band the brightest scintle falls. To investigate this issue, each observation of PSR\,J1614--2230 with a measured $\zeta$ was analyzed again, by marking the frequency of the brightest scintle in each sub-band and recalculating $\zeta$ based on the new frequency values. Comparing the new scaling index values to the original ones show a small shift towards the average value of $\zeta$, but not enough to explain the large variations. We conclude that while some of the fluctuations in $\zeta$ are likely real, most of the variations seen may disappear if wider bands were available to assure a larger number of scintles in each sub-band.

\subsection{Possibility of increased timing precision}
\label{sec:timing}
In theory, it should be possible to use the scattering delay information to correct the TOAs at each epoch before fitting the data to the timing ephemerides, and in that way increase the timing precision for each pulsar in the array. 
Here, we are interested in the difference between scattering delays included in TOAs for each pulsar and have calculated a maximum variation of the delays $\Delta \tau_{\rm d} = \tau_{\rm d;max} - \tau_{\rm d;min}$. 
To investigate which pulsars may be suitable for a correction analysis, we compare $\Delta \tau_{\rm d}$ to the median TOA uncertainty, i.e., the median precision of each TOA measurement. These values are listed in Table \ref{tab:allscatt} for each of the NANOGrav pulsars.  
Investigation of these values shows that for most of the MSPs, the measured scattering delay variation is so much smaller than the TOA uncertainties that any correction will be buried in the timing errors. This is partly because the highest scattering delays (and hence narrowest scintillation bandwidths) are unresolved in the 1.5-MHz channel data, which means that the scattering delays will show up as an additional measurement uncertainty that we cannot single out with this dataset. 

For almost all of the pulsars, the TOA precision needs to be improved by around one order of magnitude before the measured scattering delays will be of the order of the TOA uncertainties. 
Two pulsars in the list are exceptions: PSRs\,B1937+21 and J1910+1256. The scintillation bandwidths for PSR\,B1937+21 are very close to (or visibly narrower than) the channel bandwidth for many epochs, and hence the scattering delay values for this pulsar may be largely underestimated.
Therefore, to test the correction hypothesis, we used scattering delays measured for J1910+1256 only. 
We created a set of corrected arrival times by subtracting the frequency scaled scattering delays from the corresponding sub-band TOA.  
We then separately analyzed the original and corrected TOA measurements by fitting the timing parameters in the currently best ephemeris, and creating residuals from which the timing precision of each set is determined. 

When using this method to analyze the data for J1910+1256, the resulting root-mean-square residual for the corrected set did not show any improvement in comparison to the uncorrected set. 
In addition, we applied the noise budget analysis described in \cite{arz15} to the data for PSR\,J1910+1256 and found no difference in the noise properties between the uncorrected and scatter-corrected TOAs. We conclude that the variations in the scattering delays of J1910+1256 are too small compared to the TOA uncertainties to produce any improvement in timing precision when corrected for.

Another important aspect to keep in mind regarding interstellar scattering delays is the scattering contamination due to dispersion delay correction \citep{fos90, cor10,kei13}. 
Scattering is usually ignored in high-precision timing analyses, but DM variations are fitted for by correcting multi-frequency TOAs to infinite frequency at the Solar System barycenter. 
As shown in Appendix A, this effect can magnify the impact of scattering delay variations by a factor of several. 
Thus, even though the scattering delay variations in the 1500-MHz band are small, scattering may still be contributing significantly to the timing error if its effects are ignored in the DM correction procedure.

During the last few years, an increasing effort has been made to use the signal-processing tool cyclic spectroscopy (CS) to estimate pulse broadening times and scattering delays by determining the impulse response function of the ISM as well as the intrinsic pulse profile for a pulsar observation \citep{dem11,wal13}. The CS technique is a powerful tool as it allows higher frequency resolution over the observing band and hence provides measured scattering delays with much higher accuracy, compared to the filter bank approach conventionally used for pulsar observations. 
However, it is important to note that for most of the pulsars reported on here, where $\tau_{\rm d} \ll \sigma_{\rm TOA}$, the CS technique would give similar results to the dynamic spectra technique used for this paper. This is because even with higher frequency resolution, the scattering delay values would be of the same magnitude regardless of which technique is used to measure them. CS may be applicable for observations with very narrow scintillation bandwidths which normally occurs at low observing frequencies and/or for high DM pulsars, as has been shown for PSR\,B1937+21 \citep{dem11}.

\section{Conclusions}
We measure scintillation bandwidths for the NANOGrav pulsars at multiple epochs by creating dynamic spectra of wide-band GUPPI and PUPPI data. The scintillation bandwidths are then converted into scattering delays and analyzed for their variation over time. The delays are found to vary significantly.
The large fluctuations of $\Delta \nu_{\rm d}$ give modulation indices higher than expected for a Kolmogorov medium, suggesting that the density spectrum of the ISM is steeper than a Kolmogorov spectrum, or alternatively that the ISM is not well described by a thin screen in the direction of the sample pulsars. 
The mean values of the measured scattering delays are often found to be slightly higher than the corresponding values in the NE2001 electron density model. 
The wide bandwidths of the receivers have allowed for an analysis of the scaling of scattering delays with observing frequency. The scaling indices measured are in most cases smaller than that expected for a Kolmogorov medium and may also vary with time. 

The maximum variation of the scattering delays can be compared to the timing precision of each pulsar, to investigate whether the timing of a particular pulsar would benefit from correction for scattering delay variations. For most of the pulsars in the NANOGrav array, $\sigma^{\rm med}_{\rm TOA} \gg \Delta\tau_{\rm d}$, implying that the scattering delay variation has only a small effect on the measured TOAs, and a correction analysis would likely not significantly improve the timing precision, as shown by a trial analysis with PSR\,J1910+1256. 
However, even if the scattering delays in the 1500-MHz band are small, the DM correction procedure may contribute to scattering contamination of the TOAs from lower frequency bands, increasing the timing error from scattering variations by up to a factor of $\sim$20 for the NANOGrav data. 

With longer time spans and more sensitive instruments, pulsars are being timed with higher and higher precision. Over the next few years, noise budget constraints for high precision timing pulsars will be increasingly important \citep[e.g.,][]{dol14} and 
better correction for ISM contributions will improve sensitivity to gravitational waves.
Implementation of frequency-dependent modeling of the pulse profile in wide-band timing \citep[e.g.,][]{pen14} may account for part of the scattering delays discussed in this paper. An analysis of simultaneous removal of DM and scattering delays as well as inclusion of wide-band timing techniques should be carried out moving forward. 
In addition, with the development of a reliable procedure to correct for ISM delays, higher scattering pulsars may be considered as potential PTA pulsars, and would help in increasing the number of pulsars suitable for inclusion in PTAs.

\section*{Acknowledgements}
The NANOGrav project receives support from the National Science Foundation (NSF) PIRE program award number 0968296.
The National Radio Astronomy Observatory is a facility of the NSF operated under cooperative agreement by Associated Universities, Inc.
The Arecibo Observatory is operated by SRI International under a cooperative agreement with the NSF (AST-1100968), and in alliance with Ana G.  M\'{e}ndez-Universidad Metropolitana, and the Universities Space Research Association. 
NANOGrav research at UBC is funded by an NSERC Discovery Grant and Discovery Accelerator Supplement and by the Canadian Institute for Advanced Research.

{\it Author contributions.}
LL undertook the majority of the data analysis, software development and calculations, as well as generated most of the text, figures and tables. 
LL, MAM, GJ, JMC, DRS, SC, TD, MTL, TJWL, and NP all contributed to the ideas and discussion that led to the analysis carried out here. 
MAM, JMC and DRS assisted with guidance throughout the project and provided significant input about the contents of this paper. MAM and GJ provided some software and ideas that led to improvements in existing software. DRS contributed with calculations leading to the analysis presented in the Appendix. ZA, JMC, TD, MTL, TJWL, MAM, DJN and TTP commented on and suggested improvements to the manuscript text. 
This paper makes use of the NANOGrav nine-year data set \citep{arz15}. ZA, KC, PBD, TD, RDF, EF, MEG, GJ, MLJ, MTL, LL, MAM, DJN, TTP, SMR, IHS, KS, JKS, and WWZ all ran observations and analyzed timing models for this data set. Additional specific contributions to the data set are summarized in \cite{arz15}.
All authors reviewed the manuscript text and figures prior to the submission of the paper.

\bibliography{9yearscattering}{}

\begin{thebibliography}{40}
\providecommand{\natexlab}[1]{#1}
\providecommand{\url}[1]{\texttt{#1}}
\expandafter\ifx\csname urlstyle\endcsname\relax
  \providecommand{\doi}[1]{doi: #1}\else
  \providecommand{\doi}{doi: \begingroup \urlstyle{rm}\Url}\fi

\bibitem[{Armstrong} et~al.(1995){Armstrong}, {Rickett}, \& {Spangler}]{arm95}
{Armstrong}, J.~W., {Rickett}, B.~J., \& {Spangler}, S.~R., \emph{\apj}, 443,
  \penalty0 209--221, 1995.

\bibitem[{Arzoumanian} et~al.(2015){Arzoumanian}, {Brazier}, {Burke-Spolaor},
  {Chamberlin}, {Chatterjee}, {Christy}, {Cordes}, {Cornish}, {Crowter},
  {Demorest}, {Dolch}, {Ellis}, {Ferdman}, {Fonseca}, {Garver-Daniels},
  {Gonzalez}, {Jenet}, {Jones}, {Jones}, {Kaspi}, {Koop}, {Lazio}, {Lam},
  {Levin}, {Lommen}, {Lorimer}, {Luo}, {Lynch}, {Madison}, {McLaughlin},
  {McWilliams}, {Nice}, {Palliyaguru}, {Pennucci}, {Ransom}, {Siemens},
  {Stairs}, {Stinebring}, {Stovall}, {Swiggum}, {Vallisneri}, {van Haasteren},
  {Wang}, \& {Zhu}]{arz15}
{Arzoumanian}, Z., {Brazier}, A., {Burke-Spolaor}, S., {Chamberlin}, S.,
  {Chatterjee}, S., {Christy}, B., {Cordes}, J.~M., {Cornish}, N., {Crowter},
  K., {Demorest}, P.~B., {Dolch}, T., {Ellis}, J.~A., {Ferdman}, R.~D.,
  {Fonseca}, E., {Garver-Daniels}, N., {Gonzalez}, M.~E., {Jenet}, F.~A.,
  {Jones}, G., {Jones}, M., {Kaspi}, V.~M., {Koop}, M., {Lazio}, T.~J.~W.,
  {Lam}, M.~T., {Levin}, L., {Lommen}, A.~N., {Lorimer}, D.~R., {Luo}, J.,
  {Lynch}, R.~S., {Madison}, D., {McLaughlin}, M.~A., {McWilliams}, S.~T.,
  {Nice}, D.~J., {Palliyaguru}, N., {Pennucci}, T.~T., {Ransom}, S.~M.,
  {Siemens}, X., {Stairs}, I.~H., {Stinebring}, D.~R., {Stovall}, K.,
  {Swiggum}, J.~K., {Vallisneri}, M., {van Haasteren}, R., {Wang}, Y., \&
  {Zhu}, W., \emph{ArXiv e-prints}, 2015.

\bibitem[{Bhat} et~al.(1998){Bhat}, {Gupta}, \& {Rao}]{bha98}
{Bhat}, N.~D.~R., {Gupta}, Y., \& {Rao}, A.~P., \emph{\apj}, 500, \penalty0
  262, 1998.

\bibitem[{Bhat} et~al.(1999){Bhat}, {Gupta}, \& {Rao}]{bha99}
{Bhat}, N.~D.~R., {Gupta}, Y., \& {Rao}, A.~P., \emph{\apj}, 514, \penalty0
  249--271, 1999.

\bibitem[{Bhat} et~al.(2004){Bhat}, {Cordes}, {Camilo}, {Nice}, \&
  {Lorimer}]{bha04}
{Bhat}, N.~D.~R., {Cordes}, J.~M., {Camilo}, F., {Nice}, D.~J., \& {Lorimer},
  D.~R., \emph{\apj}, 605, \penalty0 759--783, 2004.

\bibitem[{Bogdanov} et~al.(2002){Bogdanov}, {Pruszy{\'n}ska}, {Lewandowski}, \&
  {Wolszczan}]{bog02}
{Bogdanov}, S., {Pruszy{\'n}ska}, M., {Lewandowski}, W., \& {Wolszczan}, A.,
  \emph{\apj}, 581, \penalty0 495--500, 2002.

\bibitem[{Champion} et~al.(2008){Champion}, {Ransom}, {Lazarus}, {Camilo},
  {Bassa}, {Kaspi}, {Nice}, {Freire}, {Stairs}, {van Leeuwen}, {Stappers},
  {Cordes}, {Hessels}, {Lorimer}, {Arzoumanian}, {Backer}, {Bhat},
  {Chatterjee}, {Cognard}, {Deneva}, {Faucher-Gigu{\`e}re}, {Gaensler}, {Han},
  {Jenet}, {Kasian}, {Kondratiev}, {Kramer}, {Lazio}, {McLaughlin},
  {Venkataraman}, \& {Vlemmings}]{cha08}
{Champion}, D.~J., {Ransom}, S.~M., {Lazarus}, P., {Camilo}, F., {Bassa}, C.,
  {Kaspi}, V.~M., {Nice}, D.~J., {Freire}, P.~C.~C., {Stairs}, I.~H., {van
  Leeuwen}, J., {Stappers}, B.~W., {Cordes}, J.~M., {Hessels}, J.~W.~T.,
  {Lorimer}, D.~R., {Arzoumanian}, Z., {Backer}, D.~C., {Bhat}, N.~D.~R.,
  {Chatterjee}, S., {Cognard}, I., {Deneva}, J.~S., {Faucher-Gigu{\`e}re},
  C.-A., {Gaensler}, B.~M., {Han}, J., {Jenet}, F.~A., {Kasian}, L.,
  {Kondratiev}, V.~I., {Kramer}, M., {Lazio}, J., {McLaughlin}, M.~A.,
  {Venkataraman}, A., \& {Vlemmings}, W., \emph{Science}, 320, \penalty0
  1309--, 2008.

\bibitem[{Chatterjee} et~al.(2009){Chatterjee}, {Brisken}, {Vlemmings}, {Goss},
  {Lazio}, {Cordes}, {Thorsett}, {Fomalont}, {Lyne}, \& {Kramer}]{cha09}
{Chatterjee}, S., {Brisken}, W.~F., {Vlemmings}, W.~H.~T., {Goss}, W.~M.,
  {Lazio}, T.~J.~W., {Cordes}, J.~M., {Thorsett}, S.~E., {Fomalont}, E.~B.,
  {Lyne}, A.~G., \& {Kramer}, M., \emph{\apj}, 698, \penalty0 250--265, 2009.

\bibitem[{Coles} et~al.(2010){Coles}, {Rickett}, {Gao}, {Hobbs}, \&
  {Verbiest}]{col10}
{Coles}, W.~A., {Rickett}, B.~J., {Gao}, J.~J., {Hobbs}, G., \& {Verbiest},
  J.~P.~W., \emph{\apj}, 717, \penalty0 1206--1221, 2010.

\bibitem[{Coles} et~al.(2015){Coles}, {Kerr}, {Shannon}, {Hobbs}, {Manchester},
  {You}, {Bailes}, {Bhat}, {Burke-Spolaor}, {Dai}, {Keith}, {Levin},
  {Os{\l}owski}, {Ravi}, {Reardon}, {Toomey}, {van Straten}, {Wang}, {Wen}, \&
  {Zhu}]{col15}
{Coles}, W.~A., {Kerr}, M., {Shannon}, R.~M., {Hobbs}, G.~B., {Manchester},
  R.~N., {You}, X.-P., {Bailes}, M., {Bhat}, N.~D.~R., {Burke-Spolaor}, S.,
  {Dai}, S., {Keith}, M.~J., {Levin}, Y., {Os{\l}owski}, S., {Ravi}, V.,
  {Reardon}, D., {Toomey}, L., {van Straten}, W., {Wang}, J.~B., {Wen}, L., \&
  {Zhu}, X.~J., \emph{\apj}, 808:\penalty0 113, 2015.

\bibitem[{Cordes}(1986)]{cor86b}
{Cordes}, J.~M., \emph{\apj}, 311, \penalty0 183--196, 1986.

\bibitem[{Cordes} \& {Lazio}(2001)]{cor01}
{Cordes}, J.~M. \& {Lazio}, T.~J.~W., \emph{\apj}, 549, \penalty0 997--1010,
  2001.

\bibitem[{Cordes} \& {Lazio}(2002)]{cor02}
{Cordes}, J.~M. \& {Lazio}, T.~J.~W., \emph{ArXiv Astrophysics e-prints}, 2002.

\bibitem[{Cordes} \& {Shannon}(2010)]{cor10}
{Cordes}, J.~M. \& {Shannon}, R.~M., \emph{ArXiv e-prints}, 2010.

\bibitem[{Cordes} et~al.(1985){Cordes}, {Weisberg}, \& {Boriakoff}]{cor85}
{Cordes}, J.~M., {Weisberg}, J.~M., \& {Boriakoff}, V., \emph{\apj}, 288,
  \penalty0 221--247, 1985.

\bibitem[{Cordes} et~al.(1990){Cordes}, {Wolszczan}, {Dewey}, {Blaskiewicz}, \&
  {Stinebring}]{cor90}
{Cordes}, J.~M., {Wolszczan}, A., {Dewey}, R.~J., {Blaskiewicz}, M., \&
  {Stinebring}, D.~R., \emph{\apj}, 349, \penalty0 245--261, 1990.

\bibitem[{Cordes} et~al.(1991){Cordes}, {Weisberg}, {Frail}, {Spangler}, \&
  {Ryan}]{cor91}
{Cordes}, J.~M., {Weisberg}, J.~M., {Frail}, D.~A., {Spangler}, S.~R., \&
  {Ryan}, M., \emph{\nat}, 354, \penalty0 121--124, 1991.

\bibitem[{Cordes} et~al.(2015){Cordes}, {Shannon}, \& {Stinebring}]{cor15}
{Cordes}, J.~M., {Shannon}, R.~M., \& {Stinebring}, D.~R., \emph{ArXiv
  e-prints}, 2015.

\bibitem[{Demorest}(2011)]{dem11}
{Demorest}, P.~B., \emph{\mnras}, 416, \penalty0 2821--2826, 2011.

\bibitem[{Demorest} et~al.(2013){Demorest}, {Ferdman}, {Gonzalez}, {Nice},
  {Ransom}, {Stairs}, {Arzoumanian}, {Brazier}, {Burke-Spolaor}, {Chamberlin},
  {Cordes}, {Ellis}, {Finn}, {Freire}, {Giampanis}, {Jenet}, {Kaspi}, {Lazio},
  {Lommen}, {McLaughlin}, {Palliyaguru}, {Perrodin}, {Shannon}, {Siemens},
  {Stinebring}, {Swiggum}, \& {Zhu}]{dem13}
{Demorest}, P.~B., {Ferdman}, R.~D., {Gonzalez}, M.~E., {Nice}, D., {Ransom},
  S., {Stairs}, I.~H., {Arzoumanian}, Z., {Brazier}, A., {Burke-Spolaor}, S.,
  {Chamberlin}, S.~J., {Cordes}, J.~M., {Ellis}, J., {Finn}, L.~S., {Freire},
  P., {Giampanis}, S., {Jenet}, F., {Kaspi}, V.~M., {Lazio}, J., {Lommen},
  A.~N., {McLaughlin}, M., {Palliyaguru}, N., {Perrodin}, D., {Shannon}, R.~M.,
  {Siemens}, X., {Stinebring}, D., {Swiggum}, J., \& {Zhu}, W.~W., \emph{\apj},
  762:\penalty0 94, 2013.

\bibitem[{Dewey} et~al.(1988){Dewey}, {Cordes}, {Wolszczan}, \&
  {Weisberg}]{dew88}
{Dewey}, R.~J., {Cordes}, J.~M., {Wolszczan}, A., \& {Weisberg}, J.~M.
\newblock \emph{{Interstellar scintillations of binary pulsars}}.
\newblock In {Cordes}, J.~M., {Rickett}, B.~J., \& {Backer}, D.~C., editors,
  \emph{Radio Wave Scattering in the Interstellar Medium}, volume 174 of
  \emph{American Institute of Physics Conference Series}, pages 217--221, 1988.

\bibitem[{Dolch} et~al.(2014){Dolch}, {Lam}, {Cordes}, {Chatterjee}, {Bassa},
  {Bhattacharyya}, {Champion}, {Cognard}, {Crowter}, {Demorest}, {Hessels},
  {Janssen}, {Jenet}, {Jones}, {Jordan}, {Karuppusamy}, {Keith}, {Kondratiev},
  {Kramer}, {Lazarus}, {Lazio}, {Lee}, {McLaughlin}, {Roy}, {Shannon},
  {Stairs}, {Stovall}, {Verbiest}, {Madison}, {Palliyaguru}, {Perrodin},
  {Ransom}, {Stappers}, {Zhu}, {Dai}, {Desvignes}, {Guillemot}, {Liu}, {Lyne},
  {Perera}, {Petroff}, {Rankin}, \& {Smits}]{dol14}
{Dolch}, T., {Lam}, M.~T., {Cordes}, J., {Chatterjee}, S., {Bassa}, C.,
  {Bhattacharyya}, B., {Champion}, D.~J., {Cognard}, I., {Crowter}, K.,
  {Demorest}, P.~B., {Hessels}, J.~W.~T., {Janssen}, G., {Jenet}, F.~A.,
  {Jones}, G., {Jordan}, C., {Karuppusamy}, R., {Keith}, M., {Kondratiev}, V.,
  {Kramer}, M., {Lazarus}, P., {Lazio}, T.~J.~W., {Lee}, K.~J., {McLaughlin},
  M.~A., {Roy}, J., {Shannon}, R.~M., {Stairs}, I., {Stovall}, K., {Verbiest},
  J.~P.~W., {Madison}, D.~R., {Palliyaguru}, N., {Perrodin}, D., {Ransom}, S.,
  {Stappers}, B., {Zhu}, W.~W., {Dai}, S., {Desvignes}, G., {Guillemot}, L.,
  {Liu}, K., {Lyne}, A., {Perera}, B.~B.~P., {Petroff}, E., {Rankin}, J.~M., \&
  {Smits}, R., \emph{\apj}, 794:\penalty0 21, 2014.

\bibitem[{Fiedler} et~al.(1987){Fiedler}, {Dennison}, {Johnston}, \&
  {Hewish}]{fie87}
{Fiedler}, R.~L., {Dennison}, B., {Johnston}, K.~J., \& {Hewish}, A.,
  \emph{\nat}, 326, \penalty0 675--678, 1987.

\bibitem[{Foster} \& {Cordes}(1990)]{fos90}
{Foster}, R.~S. \& {Cordes}, J.~M., \emph{\apj}, 364, \penalty0 123--135, 1990.

\bibitem[{Gaensler} et~al.(2008){Gaensler}, {Madsen}, {Chatterjee}, \&
  {Mao}]{gae08}
{Gaensler}, B.~M., {Madsen}, G.~J., {Chatterjee}, S., \& {Mao}, S.~A.,
  \emph{PASA}, 25, \penalty0 184--200, 2008.

\bibitem[{Gupta} et~al.(1994){Gupta}, {Rickett}, \& {Lyne}]{gup94}
{Gupta}, Y., {Rickett}, B.~J., \& {Lyne}, A.~G., \emph{\mnras}, 269, \penalty0
  1035, 1994.

\bibitem[{Hemberger} \& {Stinebring}(2008)]{hem08}
{Hemberger}, D.~A. \& {Stinebring}, D.~R., \emph{\apjl}, 674, \penalty0
  L37--L40, 2008.

\bibitem[{Johnston} et~al.(1998){Johnston}, {Nicastro}, \& {Koribalski}]{joh98}
{Johnston}, S., {Nicastro}, L., \& {Koribalski}, B., \emph{\mnras}, 297,
  \penalty0 108--116, 1998.

\bibitem[{Keith} et~al.(2013){Keith}, {Coles}, {Shannon}, {Hobbs},
  {Manchester}, {Bailes}, {Bhat}, {Burke-Spolaor}, {Champion}, {Chaudhary},
  {Hotan}, {Khoo}, {Kocz}, {Os{\l}owski}, {Ravi}, {Reynolds}, {Sarkissian},
  {van Straten}, \& {Yardley}]{kei13}
{Keith}, M.~J., {Coles}, W., {Shannon}, R.~M., {Hobbs}, G.~B., {Manchester},
  R.~N., {Bailes}, M., {Bhat}, N.~D.~R., {Burke-Spolaor}, S., {Champion},
  D.~J., {Chaudhary}, A., {Hotan}, A.~W., {Khoo}, J., {Kocz}, J.,
  {Os{\l}owski}, S., {Ravi}, V., {Reynolds}, J.~E., {Sarkissian}, J., {van
  Straten}, W., \& {Yardley}, D.~R.~B., \emph{\mnras}, 429, \penalty0
  2161--2174, 2013.

\bibitem[{Lambert} \& {Rickett}(1999)]{lam99}
{Lambert}, H.~C. \& {Rickett}, B.~J., \emph{\apj}, 517, \penalty0 299--317,
  1999.

\bibitem[{Lee} et~al.(2014){Lee}, {Bassa}, {Janssen}, {Karuppusamy}, {Kramer},
  {Liu}, {Perrodin}, {Smits}, {Stappers}, {van Haasteren}, \& {Lentati}]{lee14}
{Lee}, K.~J., {Bassa}, C.~G., {Janssen}, G.~H., {Karuppusamy}, R., {Kramer},
  M., {Liu}, K., {Perrodin}, D., {Smits}, R., {Stappers}, B.~W., {van
  Haasteren}, R., \& {Lentati}, L., \emph{ArXiv e-prints}, 2014.

\bibitem[{L{\"o}hmer} et~al.(2004){L{\"o}hmer}, {Mitra}, {Gupta}, {Kramer}, \&
  {Ahuja}]{loh04}
{L{\"o}hmer}, O., {Mitra}, D., {Gupta}, Y., {Kramer}, M., \& {Ahuja}, A.,
  \emph{\aap}, 425, \penalty0 569--575, 2004.

\bibitem[{Lyne} \& {Rickett}(1968)]{lyn68}
{Lyne}, A.~G. \& {Rickett}, B.~J., \emph{\nat}, 219, \penalty0 1339--1342,
  1968.

\bibitem[{Manchester} et~al.(2013){Manchester}, {Hobbs}, {Bailes}, {Coles},
  {van Straten}, {Keith}, {Shannon}, {Bhat}, {Brown}, {Burke-Spolaor},
  {Champion}, {Chaudhary}, {Edwards}, {Hampson}, {Hotan}, {Jameson}, {Jenet},
  {Kesteven}, {Khoo}, {Kocz}, {Maciesiak}, {Oslowski}, {Ravi}, {Reynolds},
  {Sarkissian}, {Verbiest}, {Wen}, {Wilson}, {Yardley}, {Yan}, \& {You}]{man13}
{Manchester}, R.~N., {Hobbs}, G., {Bailes}, M., {Coles}, W.~A., {van Straten},
  W., {Keith}, M.~J., {Shannon}, R.~M., {Bhat}, N.~D.~R., {Brown}, A.,
  {Burke-Spolaor}, S.~G., {Champion}, D.~J., {Chaudhary}, A., {Edwards}, R.~T.,
  {Hampson}, G., {Hotan}, A.~W., {Jameson}, A., {Jenet}, F.~A., {Kesteven},
  M.~J., {Khoo}, J., {Kocz}, J., {Maciesiak}, K., {Oslowski}, S., {Ravi}, V.,
  {Reynolds}, J.~R., {Sarkissian}, J.~M., {Verbiest}, J.~P.~W., {Wen}, Z.~L.,
  {Wilson}, W.~E., {Yardley}, D., {Yan}, W.~M., \& {You}, X.~P., \emph{\pasa},
  30:\penalty0 e017, 2013.

\bibitem[{Nicastro} et~al.(2001){Nicastro}, {Nigro}, {D'Amico}, {Lumiella}, \&
  {Johnston}]{nic01}
{Nicastro}, L., {Nigro}, F., {D'Amico}, N., {Lumiella}, V., \& {Johnston}, S.,
  \emph{\aap}, 368, \penalty0 1055--1062, 2001.

\bibitem[{Pennucci} et~al.(2014){Pennucci}, {Demorest}, \& {Ransom}]{pen14}
{Pennucci}, T.~T., {Demorest}, P.~B., \& {Ransom}, S.~M., \emph{\apj},
  790:\penalty0 93, 2014.

\bibitem[{Rickett}(1977)]{ric77}
{Rickett}, B.~J., \emph{\araa}, 15, \penalty0 479--504, 1977.

\bibitem[{Romani} et~al.(1986){Romani}, {Narayan}, \& {Blandford}]{rom86}
{Romani}, R.~W., {Narayan}, R., \& {Blandford}, R., \emph{\mnras}, 220,
  \penalty0 19--49, 1986.

\bibitem[{Scheuer}(1968)]{sch68}
{Scheuer}, P.~A.~G., \emph{\nat}, 218, \penalty0 920--922, 1968.

\bibitem[{Walker} et~al.(2013){Walker}, {Demorest}, \& {van Straten}]{wal13}
{Walker}, M.~A., {Demorest}, P.~B., \& {van Straten}, W., \emph{\apj},
  779:\penalty0 99, 2013.

\end{thebibliography}

\appendix
\section{Scattering contamination due to DM correction}
\label{app:contamination}
Most high-precision timing campaigns currently correct for DM variations, but ignore all contributions from scattering delays in the data. 
Scattering delays are then absorbed into the DM correction, resulting in systematic errors in the arrival time analysis. This is shown below for the case in which timing measurements have been made at two observing frequencies. 
Assuming that scattering scales with observing frequency as $\nu^{-4.4}$, we can write the observed TOA, $t(\nu)$,  as 
\begin{equation}
t(\nu) = t_{\infty} + D \nu^{-2} + S \nu^{-4.4}
\end{equation}
where $t_{\infty}$ is the TOA at infinite frequency, $D$ is the time dependent dispersive delay and $S$ is the time dependent scattering delay. 
By observing at two frequencies and ignoring scattering effects, we can solve for the dispersion constant $D'$ as
\begin{equation}
D' \equiv \frac{t(\nu_1) - t(\nu_2)}{\nu_1^{-2} - \nu_2^{-2}}. 
\end{equation}
However, since scattering is really present in the data, the dispersion constant will be overestimated:
\begin{align}
D' &= \frac{(t_\infty + D \nu_1^{-2} + S \nu_1^{-4.4}) - (t_\infty + D \nu_2^{-2} + S \nu_2^{-4.4})}{\nu_1^{-2} - \nu_2^{-2}} 
	\nonumber \\
    &= D + S \frac{\nu_1^{-4.4} - \nu_2^{-4.4}}{\nu_1^{-2} - \nu_2^{-2}}.
\end{align}
This overestimated $D'$ will then be used when determining $t_\infty$ from the measured $t(\nu)$. Using $t(\nu_1)$ to calculate the systematic error in arrival time at infinite frequency yields
\begin{align}
t'_{\infty}   & = t(\nu_1) - D' \nu_1^{-2} - S \nu_1^{-4.4} 
              	\nonumber \\
              	&= t_{\infty} - S \nu_1^{-2} \frac{\nu_1^{-4.4} - \nu_2^{-4.4}}{\nu_1^{-2} - \nu_2^{-2}},
\end{align}
hence the TOA will be overcorrected due to the overestimated DM and will cause an inferred infinite-frequency TOA that arrives too early with an error of 
\begin{equation}
\epsilon_{\rm TOA} = - S \nu_1^{-2} \frac{\nu_1^{-4.4} - \nu_2^{-4.4}}{\nu_1^{-2} - \nu_2^{-2}}.
\end{equation}
Finally,  
the DM correction procedure will increase the scattering variation in the $\nu_1$-band with a factor of 
\begin{equation}
\epsilon_{\rm scattering} = \frac{\epsilon_{\rm TOA}}{S \nu_1^{-4.4}} = \nu_1^{2.4} \frac{\nu_1^{-4.4} - \nu_2^{-4.4}}{\nu_1^{-2} - \nu_2^{-2}}
\end{equation}
compared to the intrinsic scattering variation in that band. For the frequency band considered in this paper, the scattering delay variation at the lowest part of the band (1100\,MHz) will therefore increase the timing error from scattering by a factor of $\sim3.4$ at the reference frequency (1500\,MHz), if scattering delays are ignored. 
For the pairs of frequency bands used in the full NANOGrav data set, the factors are 21.8 for (430\,MHz, 1500\,MHz), 5.7 for (820\,MHz, 1500\,MHz) and 3.5 for (1500\,MHz, 2100\,MHz).

\end{document}